\documentclass[trackchanges]{aastex701}

\usepackage{graphicx}
\usepackage{hyperref}

\begin{document}

\title{ZTF25abjmnps (AT2025ulz) and S250818k:  \\
A Candidate Superkilonova from a Sub-threshold Sub-Solar Gravitational Wave Trigger}

%% An Optical Transient Spatially and Temporally Coincident with a Sub-threshold Sub-Solar Gravitational Wave Trigger / Neutron Star Merger

%\collaboration{all}{The Fantabulous ZTF+GROWTH++ Collaboration}

\author[0000-0002-5619-4938]{Mansi M. Kasliwal}\email{mansi@astro.caltech.edu}
\affiliation{Division of Physics, Mathematics and Astronomy, California Institute of Technology, Pasadena, CA 91125, USA}

\author[0000-0002-2184-6430]{Tomás Ahumada}\email{tahumada@astro.caltech.edu}
\affiliation{Division of Physics, Mathematics and Astronomy, California Institute of Technology, Pasadena, CA 91125, USA}

\author[0000-0003-2434-0387]{Robert Stein}\email{rdstein@umd.edu}
\affiliation{Department of Astronomy, University of Maryland, College Park, MD 20742, USA} \affiliation{Joint Space-Science Institute, University of Maryland, College Park, MD 20742, USA} \affiliation{Astrophysics Science Division, NASA Goddard Space Flight Center, MC 661, Greenbelt, MD 20771, USA}

\author[0000-0003-2758-159X]{Viraj Karambelkar}\email{viraj@astro.caltech.edu}
\altaffiliation{NASA Hubble Fellow} \affiliation{Division of Physics, Mathematics and Astronomy, California Institute of Technology, Pasadena, CA 91125, USA} \affiliation{Columbia University, 538 West 120th Street 704, MC 5255, New York, NY 10027} 

\author[0000-0002-9364-5419]{Xander J. Hall}\email{xjh@andrew.cmu.edu}
\affiliation{McWilliams Center for Cosmology and Astrophysics, Department of Physics, Carnegie Mellon University, 5000 Forbes Avenue, Pittsburgh, PA 15213 }

\author[0000-0003-2091-622X]{Avinash Singh}\email{avinash21292@gmail.com}
\affiliation{The Oskar Klein Centre, Department of Astronomy, Stockholm University, AlbaNova, SE-10691 Stockholm, Sweden}

\author[0000-0002-4223-103X]{Christoffer Fremling}\email{fremling@caltech.edu}
\affiliation{Caltech Optical Observatories, California Institute of Technology, Pasadena, CA 91125, USA} \affiliation{Division of Physics, Mathematics and Astronomy, California Institute of Technology, Pasadena, CA 91125, USA}

\author[0000-0002-4670-7509]{Brian D.~Metzger}\email{bdm2129@columbia.edu}
\affiliation{Department of Physics and Columbia Astrophysics Laboratory, Columbia University, New York, NY 10027, USA}
\affiliation{Center for Computational Astrophysics, Flatiron Institute, 162 5th Ave, New York, NY 10010, USA}

\author[0000-0002-8255-5127]{Mattia Bulla}\email{mattia.bulla@unife.it}
\affiliation{Department of Physics and Earth Science, University of Ferrara, via Saragat 1, I-44122 Ferrara, Italy} \affiliation{INFN, Sezione di Ferrara, via Saragat 1, I-44122 Ferrara, Italy} \affiliation{INAF, Osservatorio Astronomico d’Abruzzo, via Mentore Maggini snc, 64100 Teramo, Italy}

\author[0000-0002-7942-8477]{Vishwajeet Swain}\email{vishwajeet.s@iitb.ac.in}
\affiliation{Department of Physics, Indian Institute of Technology Bombay, Powai, Mumbai 400076, India}

\author[0000-0002-7686-3334]{Sarah Antier}\email{antier@ijclab.in2p3.fr}
\affiliation{IJCLab, Univ Paris-Saclay, CNRS/IN2P3, Orsay, France}

\author[0000-0003-3224-2146]{Marion Pillas}\email{marion.pillas@iap.fr}
\affiliation{Institut d’Astrophysique de Paris, Sorbonne Université and CNRS, UMR 7095, 98 bis bd Arago, 75014 Paris, France}

\author[0009-0001-0574-2332]{Malte Busmann}\email{m.busmann@physik.lmu.de}
\affiliation{University Observatory, Faculty of Physics, Ludwig-Maximilians-Universität München, Scheinerstr. 1, 81679 Munich, Germany}

\author[0009-0006-7990-0547]{James Freeburn}\email{jamesfreeburn54@gmail.com}
\affiliation{Department of Physics and Astronomy,  University of North Carolina at Chapel Hill, Chapel Hill, NC 27599-3255, USA}

\author[0000-0003-0035-651X]{Sergey Karpov}\email{karpov@fzu.cz}
\affiliation{FZU - Institute of Physics of the Czech Academy of Sciences, Na Slovance 1999/2, CZ-182 21, Praha, Czech Republic}

\author[0009-0008-2714-2507]{Aleksandra Bochenek}\email{a.m.bochenek@2023.ljmu.ac.uk}
\affiliation{Astrophysics Research Institute, Liverpool John Moores University, 146 Brownlow Hill, Liverpool L3 5RF, UK}

\author[0000-0002-9700-0036]{Brendan O'Connor}\email{boconno2@andrew.cmu.edu}
\affiliation{McWilliams Center for Cosmology and Astrophysics, Department of Physics, Carnegie Mellon University, 5000 Forbes Avenue, Pittsburgh, PA 15213 }

\author[0000-0001-8472-1996]{Daniel A. Perley}\email{d.a.perley@ljmu.ac.uk}
\affiliation{Astrophysics Research Institute, Liverpool John Moores University, 146 Brownlow Hill, Liverpool L3 5RF, UK}

\author[0009-0006-4358-9929]{Dalya Akl}\email{dahliaaql01@gmail.com}
\affiliation{New York University Abu Dhabi, PO Box 129188, Saadiyat Island, Abu Dhabi, UAE}
\affiliation{Center for Astrophysics and Space Science (CASS), New York University Abu Dhabi, Saadiyat Island, PO Box 129188, Abu Dhabi, UAE}

\author[0000-0003-3768-7515]{Shreya Anand}\email{sanand08@stanford.edu}
\altaffiliation{LSST-DA Catalyst Postdoctoral Fellow} \affiliation{Kavli Institute for Particle Astrophysics and Cosmology, Stanford University, 452 Lomita Mall, Stanford, CA 94305, USA} \affiliation{Department of Astronomy, University of California, Berkeley, CA 94720-3411, USA}

\author[0009-0008-9546-2035]{Andrew Toivonen}\email{toivo032@umn.edu}
\affiliation{School of Physics and Astronomy, University of Minnesota, Minneapolis, Minnesota 55455, USA}

\author[0000-0003-4725-4481]{Sam Rose}\email{srose@caltech.edu}
\affiliation{Division of Physics, Mathematics and Astronomy, California Institute of Technology, Pasadena, CA 91125, USA}

\author[0009-0003-6181-4526]{Theophile Jegou du Laz}\email{tdulaz@caltech.edu}
\affiliation{Division of Physics, Mathematics and Astronomy, California Institute of Technology, Pasadena, CA 91125, USA}

\author[0000-0002-7866-4531]{Chang Liu}\email{ptg.cliu@u.northwestern.edu}
\affiliation{Department of Physics and Astronomy, Northwestern University, 2145 Sheridan Road, Evanston, IL 60208, USA}
\affiliation{Center for Interdisciplinary Exploration and Research in Astrophysics (CIERA), Northwestern University, 1800 Sherman Ave, Evanston, IL 60201, USA}

\author[0000-0001-8372-997X]{Kaustav Das}\email{kdas@astro.caltech.edu}
\affiliation{Division of Physics, Mathematics and Astronomy, California Institute of Technology, Pasadena, CA 91125, USA}

\author[0000-0003-1314-4241]{Sushant Sharma Chaudhary}\email{ssharmac@umn.edu}
\affiliation{School of Physics and Astronomy, University of Minnesota, Minneapolis, Minnesota 55455, USA}

\author[0000-0002-4843-345X]{Tyler Barna}\email{barna314@umn.edu}
\affiliation{School of Physics and Astronomy, University of Minnesota, Minneapolis, Minnesota 55455, USA}

\author[0009-0005-2987-0688]{Aditya Pawan Saikia}\email{adityaps@iitb.ac.in}
\affiliation{Department of Physics, Indian Institute of Technology Bombay, Powai, Mumbai 400076, India}

\author[0000-0002-8977-1498]{Igor Andreoni}\email{igor.andreoni@unc.edu}
\affiliation{Department of Physics and Astronomy,  University of North Carolina at Chapel Hill, Chapel Hill, NC 27599-3255, USA}

\author[0000-0001-8018-5348]{Eric C. Bellm}\email{ecbellm@uw.edu}
\affiliation{DIRAC Institute, Department of Astronomy, University of Washington, 3910 15th Avenue NE, Seattle, WA 98195, USA}

\author[0000-0002-6112-7609]{Varun Bhalerao}\email{varunb@iitb.ac.in}
\affiliation{Department of Physics, Indian Institute of Technology Bombay, Powai, Mumbai 400076, India}

\author[0000-0003-1673-970X]{S. Bradley Cenko}\email{brad.cenko@nasa.gov}
\affiliation{Astrophysics Science Division, NASA Goddard Space Flight Center, MC 661, Greenbelt, MD 20771, USA} \affiliation{Joint Space-Science Institute, University of Maryland, College Park, MD 20742, USA}

\author[0000-0002-8262-2924]{Michael W. Coughlin}\email{cough052@umn.edu}
\affiliation{School of Physics and Astronomy, University of Minnesota, Minneapolis, Minnesota 55455, USA}

\author[0000-0003-3270-7644]{Daniel Gruen}\email{daniel.gruen@lmu.de}
\affiliation{University Observatory, Faculty of Physics, Ludwig-Maximilians-Universität München, Scheinerstr. 1, 81679 Munich, Germany}\affiliation{Excellence Cluster ORIGINS, Boltzmannstr. 2, 85748 Garching, Germany}
 
\author[0000-0002-5981-1022]{Daniel Kasen}\email{kasen@berkeley.edu} 
\affiliation{Department of Physics and Department of Astronomy, University of California, Berkeley, CA 94720, USA}
\affiliation{Nuclear Science Division, Lawrence Berkeley National Laboratory, 1 Cyclotron Road, Berkeley, CA 94720, USA}

\author[0000-0001-9515-478X]{Adam A. Miller}\email{amiller@northwestern.edu}
\affiliation{Department of Physics and Astronomy, Northwestern University, 2145 Sheridan Road, Evanston, IL 60208, USA}
\affiliation{Center for Interdisciplinary Exploration and Research in Astrophysics (CIERA), Northwestern University, 1800 Sherman Ave, Evanston, IL 60201, USA}
\affiliation{NSF-Simons AI Institute for the Sky (SkAI), 172 E. Chestnut St., Chicago, IL 60611, USA}

\author[0000-0001-6573-7773]{Samaya Nissanke}\email{samaya.nissanke@desy.de}
\affiliation{Deutsches Elektronen Synchrotron DESY, Platanenallee 6, 15738 Zeuthen, Germany}
\affiliation{Deutsches Zentrum f\"ur Astrophysik (DZA), Postplatz 1, 02826 G\"orlitz, Germany}
\affiliation{Institut f{\"u}r Physik und Astronomie, Universit{\"a}t Potsdam, Haus 28, Karl-Liebknecht-Str. 24/25, 14476, Potsdam, Germany}
\affiliation{Gravitation and Astroparticle Physics Amsterdam (GRAPPA), University of Amsterdam, 1098 XH Amsterdam, The Netherlands}

\author[0000-0002-6011-0530]{Antonella Palmese}\email{palmese@cmu.edu}
\affiliation{McWilliams Center for Cosmology and Astrophysics, Department of Physics, Carnegie Mellon University, 5000 Forbes Avenue, Pittsburgh, PA 15213 }

\author[0000-0003-1546-6615]{Jesper Sollerman}\email{jesper@astro.su.se}
\affiliation{The Oskar Klein Centre, Department of Astronomy, Stockholm University, AlbaNova, SE-10691 Stockholm, Sweden}

\author{Niharika Sravan}\email{niharika.sravan@gmail.com}
\affiliation{Department of Physics, Drexel University, Philadelphia, PA 19104, USA}

\author[0000-0003-3533-7183]{G.C. Anupama}\email{gca@iiap.res.in}
\affiliation{Indian Institute of Astrophysics, 2nd Block Koramangala, 560034, Bangalore, India}

\author[0000-0001-6595-2238]{Smaranika Banerjee}\email{smaranika.banerjee@astro.su.se}
\affiliation{The Oskar Klein Centre, Department of Astronomy, Stockholm University, AlbaNova, SE-10691 Stockholm, Sweden}

\author[0000-0002-3927-5402]{Sudhanshu Barway}\email{sudhanshu.barway@iiap.res.in}
\affiliation{Indian Institute of Astrophysics, 2nd Block Koramangala, 560034, Bangalore, India}

\author[0000-0002-7777-216X]{Joshua S. Bloom}\email{joshbloom@berkeley.edu}
\affiliation{Department of Astronomy, University of California, Berkeley, CA 94720-3411, USA}

\author[0000-0002-1270-7666]{Tomás Cabrera}\email{tcabrera@andrew.cmu.edu}
\affiliation{McWilliams Center for Cosmology and Astrophysics, Department of Physics, Carnegie Mellon University, 5000 Forbes Avenue, Pittsburgh, PA 15213 }

\author[0000-0001-9152-6224]{Tracy Chen}\email{xchen@ipac.caltech.edu}
\affiliation{IPAC, California Institute of Technology, 1200 E. California Blvd, Pasadena, CA 91125, USA}

\author[0000-0001-7983-8698]{Chris Copperwheat}\email{c.m.copperwheat@ljmu.ac.uk}
\affiliation{Astrophysics Research Institute, Liverpool John Moores University, 146 Brownlow Hill, Liverpool L3 5RF, UK}

\author[0000-0001-8104-3536]{Alessandra Corsi}\email{acorsi2@jh.edu}
\affiliation{Department of Physics and Astronomy, Johns Hopkins University, 3400 N. Charles Street Baltimore, MD 21218}

\author[0000-0002-5884-7867]{Richard Dekany}\email{rgd@astro.caltech.edu}
\affiliation{Caltech Optical Observatories, California Institute of Technology, Pasadena, CA 91125, USA}

\author[0000-0001-6627-9903]{Nicholas Earley}\email{nearley@caltech.edu}
\affiliation{Division of Physics, Mathematics and Astronomy, California Institute of Technology, Pasadena, CA 91125, USA}

\author[0000-0002-3168-0139]{Matthew Graham}\email{mjg@caltech.edu}
\affiliation{Division of Physics, Mathematics and Astronomy, California Institute of Technology, Pasadena, CA 91125, USA}

\author[0009-0000-7527-205X]{Patrice Hello}\email{patrice.hello@ijclab.in2p3.fr}
\affiliation{IJCLab, Univ Paris-Saclay, CNRS/IN2P3, Orsay, France}

\author[0000-0003-3367-3415]{George Helou}\email{ghelou@caltech.edu}
\affiliation{IPAC, California Institute of Technology, 1200 E. California Blvd, Pasadena, CA 91125, USA}

\author[0000-0001-7201-1938]{Lei Hu}\email{leihu@andrew.cmu.edu}
\affiliation{McWilliams Center for Cosmology and Astrophysics, Department of Physics, Carnegie Mellon University, 5000 Forbes Avenue, Pittsburgh, PA 15213 }

\author[0000-0002-0428-8430]{Yves Kini}\email{y.kini@uva.nl}
\affiliation{Gravitation and Astroparticle Physics Amsterdam (GRAPPA), University of Amsterdam, 1098 XH Amsterdam, The Netherlands}

\author[0000-0003-2242-0244]{Ashish Mahabal}\email{aam@astro.caltech.edu}
\affiliation{Division of Physics, Mathematics and Astronomy, California Institute of Technology, Pasadena, CA 91125, USA}

\author[0000-0002-8532-9395]{Frank Masci}\email{fmasci@ipac.caltech.edu}
\affiliation{IPAC, California Institute of Technology, 1200 E. California Blvd, Pasadena, CA 91125, USA}

\author[0009-0001-4683-388X]{Tanishk Mohan}\email{tanishk.mohan@iitb.ac.in}
\affiliation{Department of Physics, Indian Institute of Technology Bombay, Powai, Mumbai 400076, India}

\author[0009-0008-8062-445X]{Natalya Pletskova}\email{np699@drexel.edu}
\affiliation{Department of Physics, Drexel University, Philadelphia, PA 19104, USA}

\author[0000-0003-1227-3738]{Josiah Purdum}\email{jpurdum@caltech.edu}
\affiliation{Caltech Optical Observatories, California Institute of Technology, Pasadena, CA  91125}

\author[0000-0003-3658-6026]{Yu-Jing Qin}\email{yujingq@caltech.edu}
\affiliation{Division of Physics, Mathematics and Astronomy, California Institute of Technology, Pasadena, CA 91125, USA}

\author[0000-0002-5683-2389]{Nabeel Rehemtulla}\email{nabeelrehemtulla2027@u.northwestern.edu}
\affiliation{Department of Physics and Astronomy, Northwestern University, 2145 Sheridan Road, Evanston, IL 60208, USA}      
\affiliation{Center for Interdisciplinary Exploration and Research in Astrophysics (CIERA), Northwestern University, 1800 Sherman Ave, Evanston, IL 60201, USA}    
\affiliation{NSF-Simons AI Institute for the Sky (SkAI), 172 E. Chestnut St., Chicago, IL 60611, USA}

\author[0000-0003-3173-4691]{Anirudh Salgundi}\email{anirudhs@unc.edu}
\affiliation{Department of Physics and Astronomy,  University of North Carolina at Chapel Hill, Chapel Hill, NC 27599-3255, USA}

\author{Yuankun Wang}\email{ykwang@uw.edu}
\affiliation{DIRAC Institute, Department of Astronomy, University of Washington, 3910 15th Avenue NE, Seattle, WA 98195, USA}

\begin{abstract}
On August 18, 2025, the LIGO-Virgo-KAGRA collaboration reported gravitational waves from a sub-threshold binary neutron star merger. If astrophysical, this event would have a surprisingly low chirp mass, suggesting that at least one neutron star was below a solar mass. The Zwicky Transient Facility mapped the coarse localization and discovered a transient, ZTF\,25abjmnps (AT2025ulz), that was spatially and temporally coincident with the gravitational wave trigger. The first week of follow-up suggested properties reminiscent of a GW170817-like kilonova. Subsequent follow-up suggests properties most similar to a young, stripped-envelope, Type IIb supernova.  Although we cannot statistically rule out chance coincidence, we undertake due diligence analysis to explore the possible association between ZTF\,25abjmnps and S250818k. Theoretical models have been proposed wherein sub-solar neutron star(s) may form (and subsequently merge) via accretion disk fragmentation or core fission inside a core-collapse supernova i.e. a ``superkilonova". Here, 
we qualitatively discuss our multi-wavelength dataset in the context of the superkilonova picture. Future higher significance gravitational wave detections of sub-solar neutron star mergers with extensive electromagnetic follow-up would conclusively resolve this tantalizing multi-messenger association.
\end{abstract}

\section{Introduction}
Multi-messenger astrophysics is the study of sources that are detected by at least two of four independent information messengers: electromagnetic radiation, gravitational waves, neutrinos, and cosmic rays. Until recently, our Sun and Supernova 1987A were the only two multi-messenger sources from which both neutrinos and photons had been extensively studied. This past decade has seen tremendous progress in the study of multi-messenger sources. The discovery of GW170817 \citep{GW170817}, a binary neutron star merger that was detected by both gravitational wave interferometers and telescopes across the entire electromagnetic spectrum, was a major breakthrough for the multi-messenger field (see \citealt{GW170817MMA} and references therein including \citealt{Coulter2017,Hallinan2017,Kasliwal2017,Evans2017,Pian2017,Tanvir2017,Drout2017,Smartt2017,SoaresSantos2017,Kasen2017,Margutti2017,Troja2017}). In the last five years, multiple compelling electromagnetic counterparts have also been proposed to high-energy neutrinos: our own galaxy \citep{ic_gp_23}, active galactic nuclei \citep{kadler_16,ic_txs_18,plavin_20,ic_ngc1068_22,ic_agn_cores_22}, tidal disruption flares \citep{stein_21,reusch_22}, and interacting hydrogen-poor supernovae \citep{stein_25}. Separately, candidate counterparts to the highest mass binary black hole mergers have also been proposed \citep{Graham2020, Graham2023, Cabrera2024}. Extensive searches in coarsely localized gravitational wave events of neutron star mergers have also led to more constraints on the nature of their electromagnetic counterparts (e.g., \citealt{Kasliwal2020, Andreoni2020, Paterson2021, dejaeger2022, Ahumada2024, Ahumada2025, Hu2025}).   

As we celebrate a decade since the discovery of the first gravitational waves \citep{GW150914}, we acknowledge that gravitational wave events have opened our eyes to categories of sources that we did not even know existed. For example, no neutron star black hole binary has yet been detected electromagnetically and now over half a dozen have been detected in gravitational waves \citep{GWTC4}. Many facets of the black hole mass function have also come as a surprise, such as black holes in the upper mass gap and lower mass gap \citep{GWTC4}. Now, the discovery of S250818k \citep{GCN41437}, the most plausible candidate to-date for merger of sub-solar mass neutron stars could be another eye-opening discovery: sub-solar compact objects could either be the long-awaited proof-of-existence of primordial black holes \citep{Zeldovich:1967lct, 1971MNRAS.152...75H, 1975ApJ...201....1C, Chapline:1975ojl} or the first evidence of an entirely new stellar evolutionary pathway that could create such small neutron stars \citep{PiroPfahl2007,Metzger2024}.  

%It is theorized these subsolar-mass objects could exist as primordial black holes, created by the collapse of overdensities during the early universe  or through the collapse of dark matter halos \citep{DAmico:2017lqj, Shandera:2018xkn, Choquette:2018lvq}.

%Previous searches of sub-solar neutron star mergers have constrained rates and a new template bank was implemented earlier this year. The most compelling (lowest FAR) sub-solar mass merger to date is S250818k. 

In this paper, we present the discovery of an optical transient, ZTF\,25abjmnps (AT2025ulz), and discuss the possibility of association with the gravitational wave trigger S250818k. In \S~\ref{sec:discovery}, we describe how the Zwicky Transient Facility (ZTF) promptly mapped the localization of the gravitational wave trigger S250818k and identified the transient ZTF\,25abjmnps as a candidate counterpart amidst other unrelated transients. In \S~\ref{sec:followup}, we present the extensive follow-up data taken by over a dozen telescopes worldwide in the optical and infrared wavelengths. In \S~\ref{sec:kilonova}, we analyze the data in the context of kilonova and afterglow models. In \S~\ref{sec:supernova}, we analyze the data in the context of supernova models and the literature sample of supernovae. In \S~\ref{sec:discussion}, 
we examine whether the observed properties of AT2025ulz could be explained in the context of a theoretical model involving core fission or fragmentation in the disk of a core-collapse supernova leading to the formation of sub-solar neutron stars that subsequently merge, i.e. a superkilonova. The term ``superkilonova" was first used to describe a theoretical model wherein a collapsar produces several solar masses of heavy elements by r-process nucleosynthesis \citep{Siegel2022}. Here, we propose to expand and generalize the use of the term superkilonova to more broadly include any core-collapse supernova that has any kilonova-like r-process nucleosynthesis inside it (e.g., accretion disk fragmentation, fissioning of core, etc.). We conclude with forward-looking next steps on what could more conclusively prove or disprove the multi-messenger association between such electromagnetic transients and gravitational wave events. 

\section{Discovery}
\label{sec:discovery}

On 2025-08-18 01:20:06.030 UTC, the LIGO/Virgo/KAGRA collaboration alerted the community \citep{Chaudhary2024} on a compact binary merger candidate S250818k during real-time processing of data from LIGO Hanford Observatory (H1), LIGO Livingston Observatory (L1), and Virgo Observatory (V1) \citep{GCN41437}. The low latency false alarm rate (FAR) was 2.1 per year, a factor of just two higher than our nominal threshold of 1 per year to automatically trigger follow-up. The low latency estimate of the terrestrial probability was 70\%. While this may seem high, we caution that low latency estimates may improve significantly with offline analysis --- specifically, the binary neutron star merger S231109ci had a low latency FAR estimate of 13 per year and terrestrial probability of 96\%, and now has a published offline FAR estimate of 1 per 50 years and is now confirmed to be a binary neutron star \citep{Niu2025}. We also refer the reader to the analysis presented in \citealt{Gillanders2025} (see their Figure 1), based on \citealt{Nicholl2025}, that shows the parameter space in which S250818k is closer to the locus of astrophysical GW triggers than terrestrial GW triggers.   

If astrophysical in origin, S250818k is the only gravitational wave candidate event from online searches that has accompanying rapid parameter estimates in the sub-solar mass regime. The binned chirp mass estimate is 0.87 $\mathrm{M}_\odot$ (highest probability) and at least one of the components is less than a solar mass at greater than 99\% confidence \citep{GCN41437}. We note that previous searches of sub-solar neutron star mergers have constrained rates \citep{LIGOScientific:2021job, LVK:2022ydq}. Targeted, online searches for gravitational waves emitted from subsolar-mass are now underway, implemented in January of 2025, by the GstLAL and MBTA SSM search algorithms \citep{Hanna:2024tom, Allene:2025saz}. To date, S250818k is the lowest FAR candidate sub-solar neutron star merger event.

\begin{figure}[hbt!]
\includegraphics[width=\textwidth]{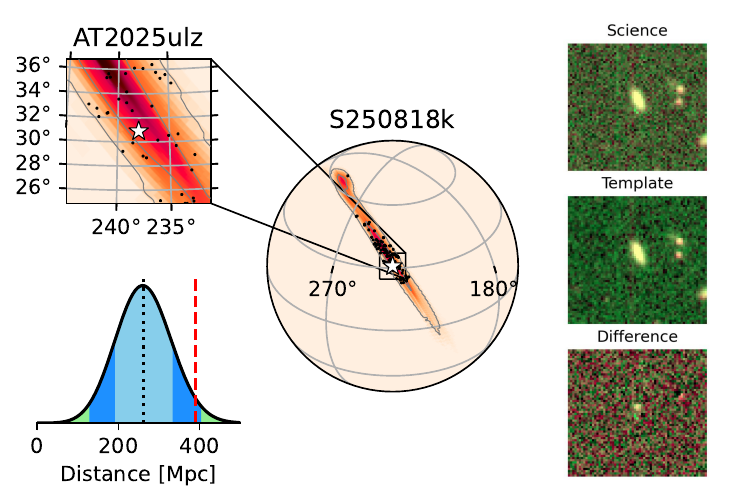}
\caption{The discovery of ZTF\,25abjmnps. \textbf{Upper Left:} Zoomed-in skymap of S250818k, with the 50\% and 95\% contours marked by grey lines, and the position of ZTF\,25abjmnps marked by a white star. The other excluded ZTF candidates are marked by black dots. ZTF\,25abjmnps lies within the central 50\% region, close to the center. \textbf{Center:} Zoomed-out skymap of S250818k. \textbf{Lower Left:} Line-of-sight distance at the position of ZTF\,25abjmnps. The median GW distance (262 Mpc) at the location of ZTF\,25abjmnps is illustrated with the vertical dotted black line, and the shaded regions correspond to the $\pm1\sigma$, $\pm2\sigma$ and $\pm3\sigma$ regions. The luminosity distance of ZTF\,25abjmnps is illustrated with the dashed red line, and lies within $2\sigma$ of the median value.
\textbf{Right:} False-color ZTF discovery image of ZTF\,25abjmnps in $g-$band and $r-$band (top), alongside the template reference image (center) and difference image (bottom). Differencing yields a clean PSF-like excess in both filters at the location of the transient.
%{\it Figure Credit: Robert}
}
\label{fig:skymap}
\end{figure}

When alerted by the \texttt{Fritz Marshal} instance of \texttt{Skyportal} \citep{skyportal_19,skyportal_23},
we deliberated on the points above and decided to manually trigger dedicated ToO observations of S250818k with ZTF \citep{ztf_bellm_19,ztf_graham_19,ztf_dekany_19} using \texttt{snipergw}\footnote{\url{https://github.com/robertdstein/snipergw}}. An observing plan was generated with \texttt{gwemopt} \citep{gwemopt_a,gwemopt_b,gwemopt_c} for the ZTF field grid, balancing enclosed probability and observability for each individual ZTF field. We observed each field with three 300s exposures (in $g$-band, $r$-band and then $g$-band), to measure both color and fade rate for all candidates. Our ZTF observations began promptly at 2025-08-18 04:02 UTC, approximately 2.7 hours after merger. We ultimately covered 33.1\% of the reported localization region at least twice within the first 36 hours of merger (369.7 sq deg, correcting for chip gaps). With 300s exposures, we reached a median depth of 21.85 (21.99) in $g-$band ($r-$band). Our images were processed using the standard ZTF data processing pipeline \citep{ztf_masci_19}, yielding individual transient alerts which were distributed to ZTF Partnership brokers. The alerts were analyzed with \texttt{nuztf}, a software package developed to identify electromagnetic counterparts to neutrinos, gravitational waves and gamma-ray bursts \citep{stein_23,nuztf}. 

With the first night of data, \texttt{nuztf} selected 58 candidates with at least two detections and no detection history prior to merger. Of these candidates, ZTF\,25abjmnps was immediately identified as the only candidate which appeared to be red, and with a host galaxy photometric redshift consistent with the estimated merger distance (see Figure \ref{fig:skymap}). The transient was reported to TNS and assigned the name AT2025ulz \citep{tns_disc}, and was also highlighted via a GCN Circular \citep{GCN41414}.
We repeated this analysis on subsequent nights. We also used independent search algorithms on ZTF data, such as ZTFReST \citep{Andreoni2021} and \texttt{Fritz Marshal} filtering, to cross-validate our candidate list. Considering only sources detected within 72 hours of merger, there were a total 30061 individual ZTF alerts in the 95\% contour. Of these, 109 candidates were manually vetted using the \texttt{Fritz Marshal}. We systematically ruled out all of the other candidates as either unrelated variable sources, sources too distant to be associated with S250818k, or as sources with too slow photometric evolution. In some cases, additional photometry was obtained in order to confirm slow photometric evolution. The full list is given in Appendix Table \ref{tab:candidates}, alongside the rejection criterion for each candidate. Apart from ZTF\,25abjmnps, no other plausible counterparts were found in our data. 
An independent mapping and search by the Pan-Starrs1 survey 
%also notes that ZTF\,25abjmnps is the only plausible counterpart and 
reports no other candidate counterpart to S250818k \citep{GCN41493,Gillanders2025}. 
In Appendix B, we discuss some candidates discussed in \citealt{Gillanders2025} and \citealt{Franz2025} as comparably compelling (or even higher ranked by their metric) to ZTF\,25abjmnps -- our conclusion remains that ZTF\,25abjmnps is the only plausible candidate counterpart to S250818k. 

% Discovered at r=21.3\,mag in a galaxy with no archival redshift, we promptly reported ZTF\,25abjmnps (AT2025ulz) as unusual and encouraged follow-up to classify this source .

\begin{figure}[hbt!]
\centering
\includegraphics[width=0.8\textwidth]{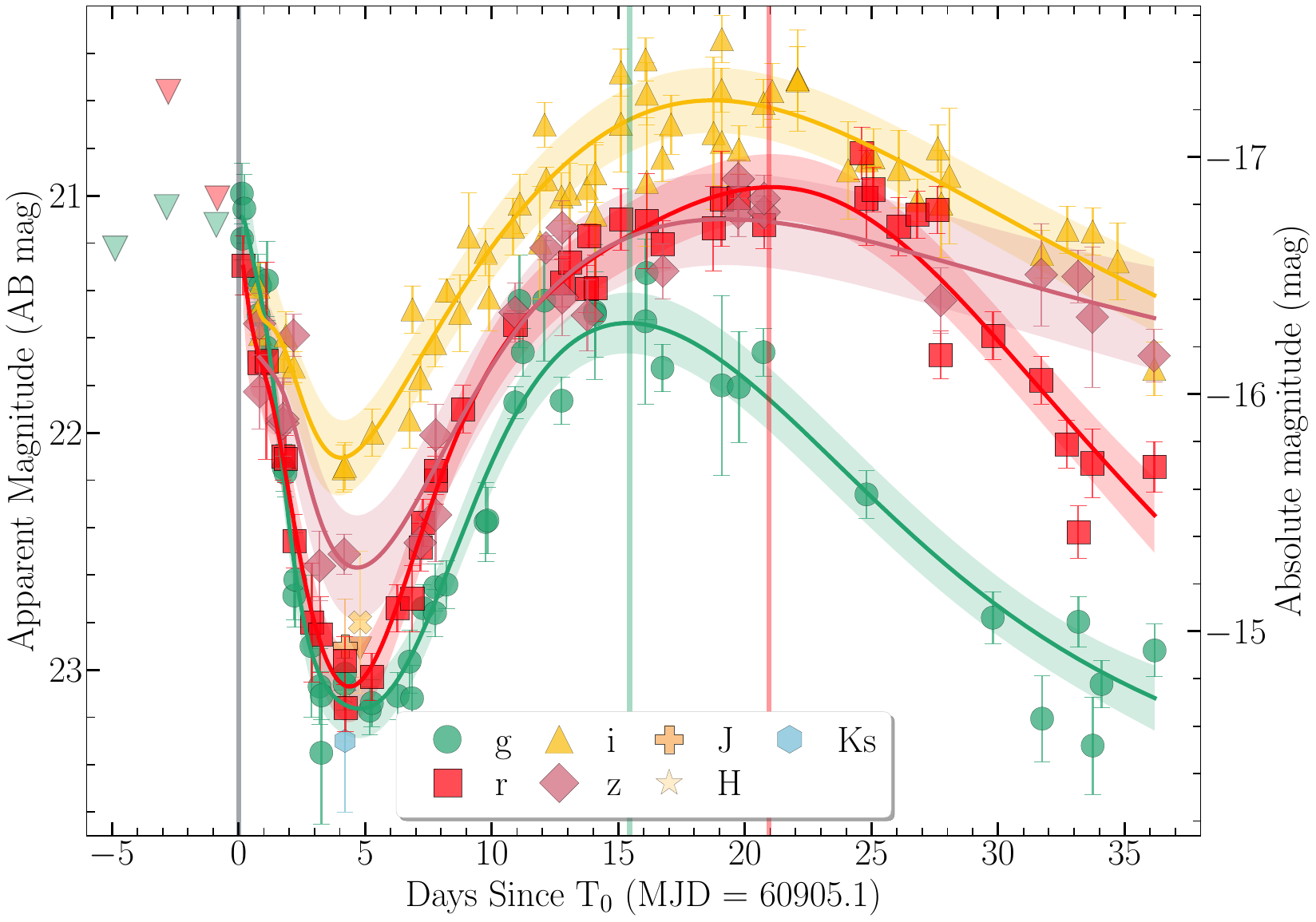}
\caption{Optical and near-infrared lightcurve of ZTF\,25abjmnps (AT2025ulz). The filters $g$, $r$, $i$, $z$, $J$,$H$ and $Ks$ are shown with different colors and different symbols. We show with a vertical line the onset of S250818k defined as $t=0$ days, another vertical line for the second g-band peak and the second r-band peak. GP fit to the light curve is shown with the solid colored lines and uncertainty as shaded region. 
%{\it Figure Credit: Tomas and Avinash}
}
\label{fig:lc}
\end{figure}

\begin{figure}[hbt!]
\includegraphics[width=\textwidth]{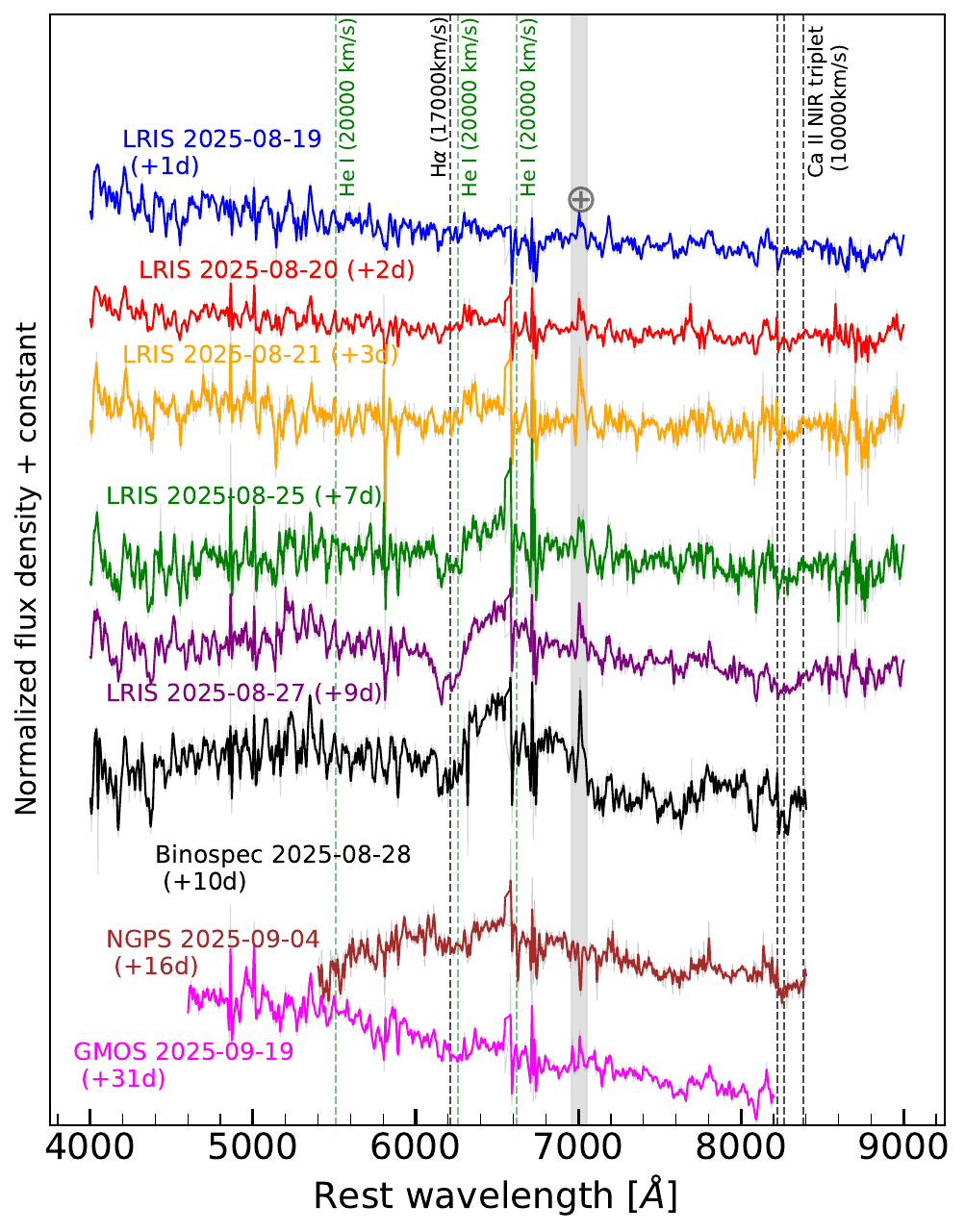} 
\caption{Spectroscopic evolution of ZTF\,25abjmnps subtracting a scaled archival host spectrum \citep{Hall2025desi,hall_at2025ulz_2025}. A blue featureless continuum evolved to become redder and then showed prominent broad P-Cygni features. Hydrogen, Helium and Calcium lines are maked with vertical dashed lines at the velocity indicated in parantheses. The spectral range impacted by telluric correction is shown with a gray shaded line and should be ignored. The flux calibration in the GMOS spectrum is impacted by using a flux calibrator on a different night instead of the same night. 
%{\it Figure Credit: Viraj and Xander}
}
\label{fig:lris}
\end{figure}

\section{Follow-up}
\label{sec:followup}
Fortuitously, ZTF\,25abjmnps was located in the overlap region between two ZTF fields, yielding a total of six ZTF observations in the first night. After the identification of ZTF\,25abjmnps as a candidate counterpart, extensive photometric follow-up was undertaken by the Fraunhofer Telescope at Wendelstein Observatory (FTW), Liverpool Telescope (LT), Canada France Hawaii Telescope (CFHT), Gemini Observatory, the two meter twins telescopes (TTT), the 2m-Himalayan Chandra Telescope (HCT), Palomar 60-in/SEDM,  and Keck I/LRIS in the optical and Keck I/MOSFIRE, P200/WIRC, FTW, ARC 3.5m/NICFPS  and WINTER in the infrared. For spectroscopic follow-up, we used Keck I/LRIS, Gemini North/GMOS and MMT/Binospec. Details about the follow-up telescopes and data analysis are in the Appendix. The follow-up effort was coordinated via the \texttt{Fritz Marshal} instance of \texttt{Skyportal}.  

To carefully cross-calibrate the photometry self-consistently across multiple telescopes, we took the following steps. For subtraction, all telescopes used the same set of templates: Legacy Survey DR10 images~\citep{2019AJ....157..168D} for $g$-band and $z$-band, the PS1 image in $i$-band, and archival MegaCam \href{https://www.skysurvey.cc/releases/}{UNIONS} survey images in $r$-band. For zeropoint calculations, all telescopes used a common set of calibration stars from the PS1 catalog. As the object is located on top of a bright host galaxy, and the passbands of the science images and templates are not exactly the same, the galaxy is not always cleanly subtracted leading to subtraction residuals that might impact the photometry. To account for this, we reduced the aperture size to reduce the contribution from biased background estimation inside it. Our error estimate includes both statistical error and a systematic error floor (0.1\,mag minimum when fainter than 21.5 mag;  0.05\,mag minimum when brighter than 21.5 mag). The systematic error floor captures the measurement variations when changing the aperture size and roughly corresponds to the scatter due to filter mismatch. A list of all photometry is given in Table \ref{tab:photometry} and shown in Figure~\ref{fig:lc}. To obtain smooth interpolations of the light curves, we used Gaussian Process (GP) regression as implemented in \texttt{scikit-learn}. %\citep{scikit-learn}.  
The GP was applied in logarithmic time space with a Matérn ($\nu=2.5$) kernel, which provides a flexible yet smooth representation of the light curve evolution while naturally incorporating photometric uncertainties. The GP fit is also shown in Figure~\ref{fig:lc}.

To correct for reddening, we consider both Galactic and intrinsic host extinction. We estimate and correct for Galactic extinction using dustmaps from \cite{SchaflyFinkbeiner2011}. The presence of narrow Na\,I D lines from the host galaxy in the spectrum can be a good proxy for host galaxy reddening \citep{Poznanski2012, Stritzinger2018}. Specifically, the equivalent width of the Na I D$_1$ and D$_2$ lines can be related to the extinction by the following empirical relation: $A\mathrm{_{v}^{host}}=0.78(\pm0.15)\,\mathrm{EW_{Na\,I\,D}}$ \citep{Stritzinger2018}. Using the LRIS spectrum taken on UT 2025-08-25, we measure $\mathrm{EW_{Na\,I\,D}}$ and calculate E(B-V)$\approx0.29$\,mag and $A\mathrm{_{v}^{host}}\approx 0.89$\,mag, adopting the coefficients from \cite{SchaflyFinkbeiner2011} (i.e. R$_\mathrm{v}=3.1$), who assumed a reddening law from \cite{Fitzpatrick1999}. For this, we opt to use a spectrum where the object is placed in the slit, rather than the center of the host galaxy, to probe the extinction of the transient's local environment. We caution that since low-resolution spectra are being used, this estimate is best used as an upper limit on the extinction \citep{Poznanski2012}.

In order to derive a redshift from the spectra, we use multiple host galaxy emission lines that yield z=0.0848 or 399.4\,Mpc, which we adopt throughout. To correct for host galaxy contamination in the transient spectra, we compare two approaches: scale and subtract an archival host galaxy spectrum vs. model the host galaxy light using the method described in \citet{LiuMiller2025}. We find using an archival host spectrum from the DESI survey \citep{Hall2025desi,hall_at2025ulz_2025}to yield overall less noisy results and show these in the spectral collage in Figure~\ref{fig:lris}. In a few months, we plan to get a host galaxy spectrum exactly at the transient location after the transient has faded and expect that to yield the cleanest results. Host galaxy light subtraction is particularly challenging as the transient is located on top of an active star forming region in the galaxy. Therefore, to facilitate comparison to supernova templates, we normalized each spectrum and the template set to a common scale. The observed spectrum is first dereddened, and the narrow host lines and telluric bands are masked. A smooth continuum is then estimated using a cubic spline using \textit{scipy}, iteratively reweighted with Tukey biweights implemented in \textit{numpy} following the standard bisquare form, which suppress the influence of strong lines and outliers and provides a stable approximation to the underlying spectral shape. The spectrum is divided by this continuum, and the resulting ratio is centered and rescaled by subtracting its median value and linearly mapping the central flux distribution to the $\pm$1 interval. This normalization emphasizes line morphology while minimizing sensitivity to absolute flux calibration (Figure~\ref{fig:sntemplates}).

\section{Is ZTF\,25abjmnps a kilonova?}
\label{sec:kilonova}
In the first few days after discovery, the fast decline, the reddening of the color, and the featureless optical spectra of ZTF\,25abjmnps were reminiscent of GW170817. A direct comparison to GW170817 as well as model fits to the light curve using a kilonova model grid gave reasonable ejecta properties. When the light curve started to plateau and then rise, we explored the hypothesis that an off-axis afterglow-like relativistic component may be contributing to the emission.
%Furthermore, kilonovae could also have two peaks based on the underlying ejecta distribution. However, no kilonova model that is based strictly on radioactive decay of heavy elements can explain the full light curve as the second peak of emission is much too long lasting and luminous. 
We find that a joint kilonova and afterglow models can indeed fit the entire optical light curve, but the derived parameters are not consistent with the non-detections in the radio and X-ray. 

%\subsection{Fitting a kilonova model}\label{sec:kilonova_component}
%In the first few nights, the photometric evolution was characterized by a rapid g-band decay and a progressive reddening of the source, a behavior reminiscent of kilonovae \citep{Metzger2019} and even resembling GW170817 (see left panels of Fig.~\ref{fig:kilo-grb-lc}). At $\sim3$~days, $g$- and $r$-band light curves began to flatten, suggesting the contribution of an additional component (see Section~\ref{sec:afterglow}). 
First, we focus on the optical data in the first 2.5~days and fit these data with kilonova models. Specifically, we employ a grid of merging binary neutron star kilonova models computed with the Monte Carlo radiative transfer code \textsc{possis} \citep{Bulla2019,Bulla2023} and recently presented in \cite{Ahumada2025}. Briefly, the grid is constructed by varying the mass, averaged velocity and averaged electron fraction of two ejecta components \citep{Nakar2020}: a first axially-symmetric high-velocity component ejected during the merger (dynamical ejecta) and a second spherically-symmetric low-velocity component ejected from the post-merger disk (disk-wind ejecta). When varying the six free parameters ($m_{\rm dyn}$, $\bar{v}_{\rm dyn}$, $\bar{Y_e}_{\rm dyn}$, $m_{\rm wind}$, $\bar{v}_{\rm wind}$, $\bar{Y_e}_{\rm wind}$) within ranges predicted by numerical relativity simulations \citep{Radice2018,Nedora2021}, and accounting for $11$ different viewing angles $\theta_{\rm obs}$, a total of $33\,792$ different kilonova models are fitted to the available photometry.

The best-fit kilonova model gave a good fit to the optical photometry data in the first 3 days (reduced $\chi^2=1.05$). The best-fit model corresponds to the following parameters: $m_{\rm dyn}=0.02\,M_\odot$, $\bar{v}_{\rm dyn}=0.2$c, $\bar{Y_e}_{\rm dyn}=0.2$, $m_{\rm wind}=0.09\,M_\odot$, $\bar{v}_{\rm wind}=0.03$c, $\bar{Y_e}_{\rm wind}=0.3$, $\theta_{\rm obs}=25.8^\circ$. In particular, the relatively high masses inferred for both ejecta components are due to the relatively high luminosities of ZTF\,25abjmnps, which is $\sim0.5-1.0$\,mag brighter than GW170817 and peaks at an absolute magnitude of $\sim-17$\,mag. An \textsc{NMMA} \citep{Pang2023} fit to the early light curve is presented in a companion paper \citep{Hall2025sn}. The inferred viewing angle and the lack of high-energy detections \citep{GCN41441} is suggestive of an off-axis event but not well-constrained by the early data alone.
%at an inclination similar to the one inferred for GW170817 \citep{Mooley2022}. 

%\subsection{Adding an afterglow component to the kilonova model}
%\label{sec:afterglow}
%Beyond 2.5~days, the light curve begins to rise again, deviating from the kilonova model grid described above. Next we explore an off-axis afterglow as a possible explanation for this re-brightening.
The peak time of an afterglow depends on the off-axis angle: the difference between the jet core angle and the viewing angle \( \theta_c - \theta_{\rm obs} \). In case of AT2017gfo, light curve modeling and VLBI measurements of the super-luminal motion of its jet yielded a viewing angle of $\approx 19\degr$, and the jet core angle was estimated to be $1.5\degr -4\degr$. The large difference led to the afterglow lightcurve peaking about 150 days after the merger \citep{2018Natur.561..355M,2019Sci...363..968G,2022Natur.610..273M,Govreen2023}. 
%If ZTF\,25abjmnps is indeed the counterpart of the compact binary merger, the lack of high-energy detections \citep{GCN41441} suggests the possibility that we may be viewing it off-axis. 
We modeled the afterglow using \textsc{JETSIMPY} \citep{2024ApJS..273...17W}, which models the data as synchrotron emission from a structured jet that interacts with an external medium. We assume a structured jet with a Gaussian profile defined as
\begin{eqnarray}
E(\theta) & = & E_{\mathrm{K,iso}} ~exp \left[ -\frac{1}{2} \left( \frac{\theta}{\theta_c} \right)^2 \right], \label{eq:gaussian_jet}\\
\Gamma(\theta) & = & (\Gamma_{0} - 1) ~exp \left[ -\frac{1}{2} \left( \frac{\theta}{\theta_c} \right)^2 \right] + 1, \label{eq:gaussian_jet2}
\end{eqnarray}
where \(E_{\mathrm{K,iso}}\) is the isotropic-equivalent energy, \( \Gamma_0 \) is the initial Lorentz factor, \( \theta_c \) is the jet half-opening angle.

\begin{figure}[hbt!]
\includegraphics[width=0.28\textwidth]{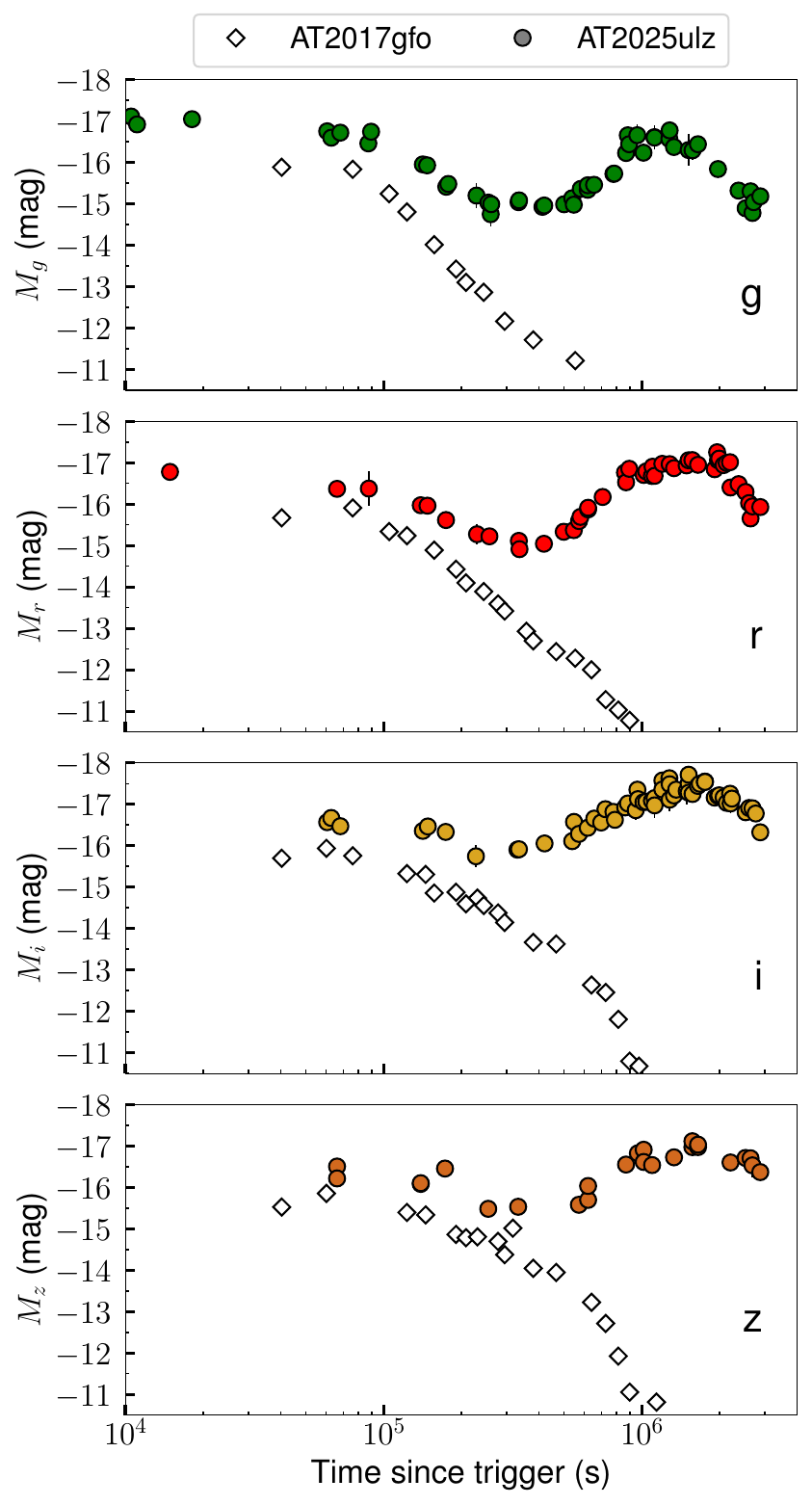}
\includegraphics[width=0.72\textwidth]{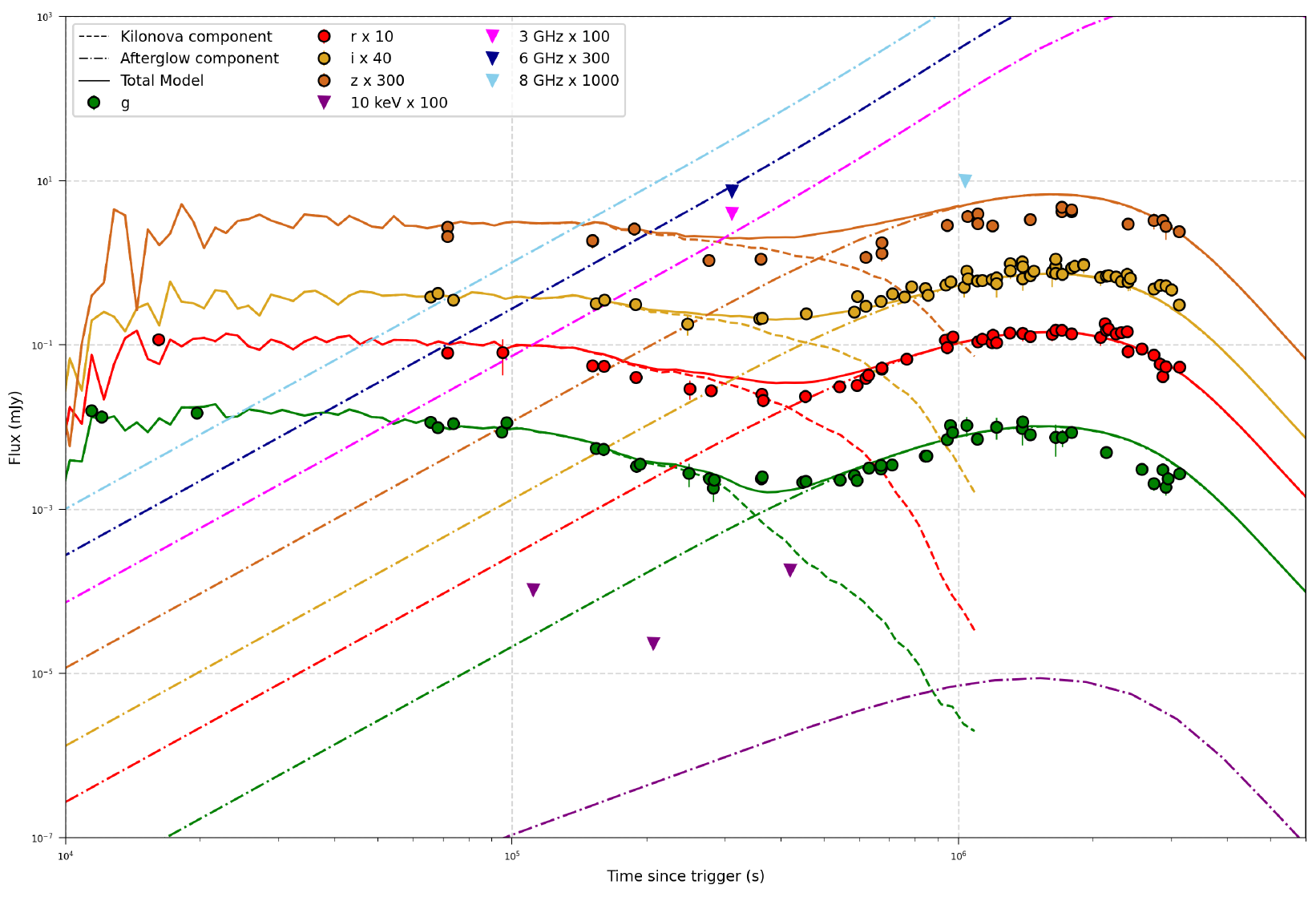}
\caption{{\bf Left panels}: $griz$ light curves of ZTF\,25abjmnps (filled circles) compared to GW170817/AT2017gfo (open diamonds). {\bf Right panel}: Combined kilonova and off-axis afterglow model plotted with observed data. The best-fit kilonova model corresponds to the following parameters: $m_{\rm dyn}=0.02\,M_\odot$, $\bar{v}_{\rm dyn}=0.2$c, $\bar{Y_e}_{\rm dyn}=0.2$, $m_{\rm wind}=0.09\,M_\odot$, $\bar{v}_{\rm wind}=0.03$c, $\bar{Y_e}_{\rm wind}=0.3$, $\theta_{\rm obs}=12^\circ$.
The afterglow fit yields \( \log_{10}(E_0 /\mathrm{erg}) = 53.88 \pm 0.08 \) with external medium density of \( \log_{10}(n_0 /\mathrm{cm}^{-3}) = -0.97 \pm 0.09\). The structured jet has initial Lorentz factor of $48.20 \pm 15.11$, with the jet core angle \( \theta_c = 1.42\degr \pm 0.12\degr\), and the viewing angle \( \theta_{\rm obs} = 12.04\degr \pm 0.21\degr\). The electron energy distribution is characterized by an index of $p = 2.88 \pm 0.08$. Note that for a better constrained fit, we fixed the microphysical parameters to typical afterglow values, adopting $\epsilon_e = 0.1$ and $\epsilon_b = 0.01$.
%{\it Figure Credit: Mattia and Vishwajeet}
} 
\label{fig:kilo-grb-lc}
\end{figure}

\begin{figure}[hbt!]
\includegraphics[width=0.5\textwidth]{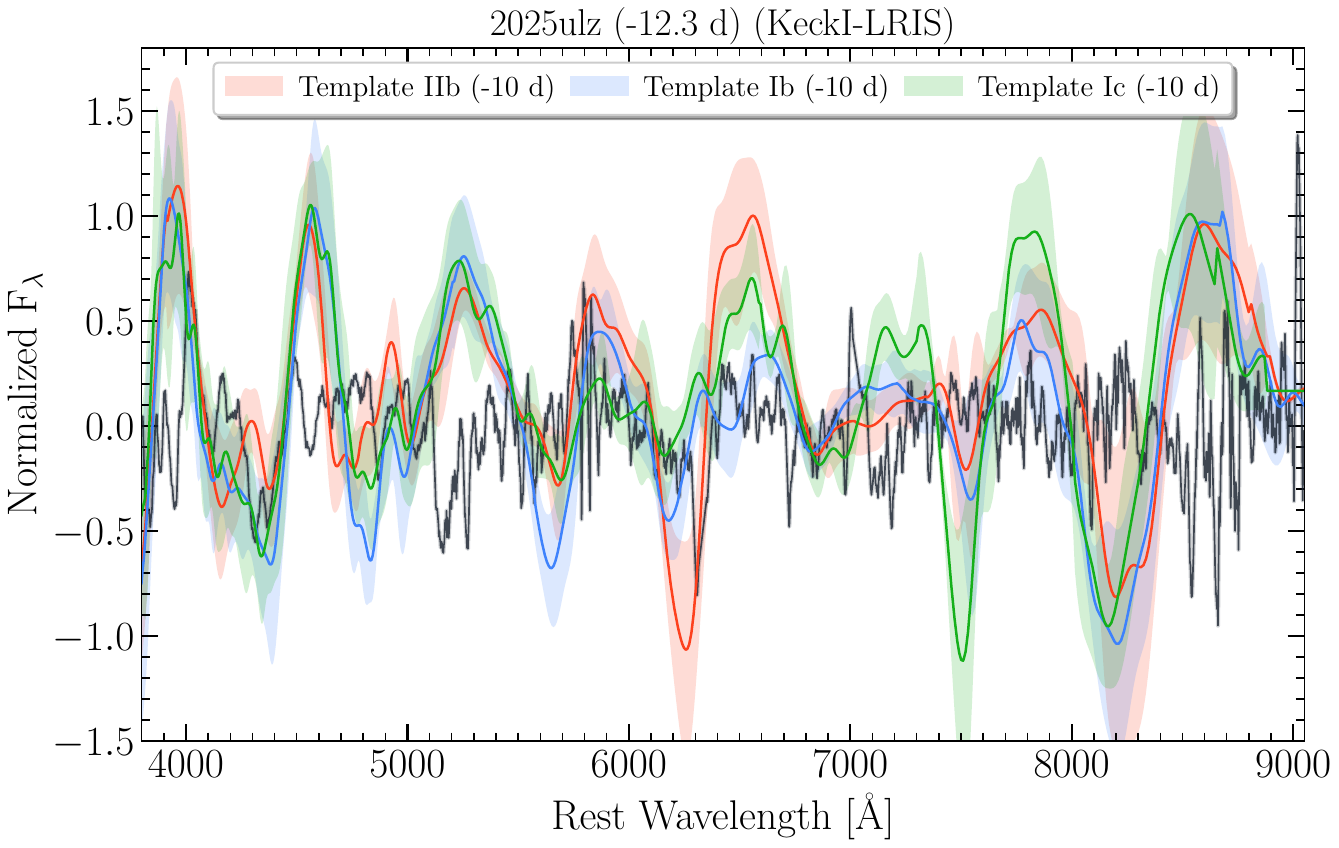}
\includegraphics[width=0.5\textwidth]{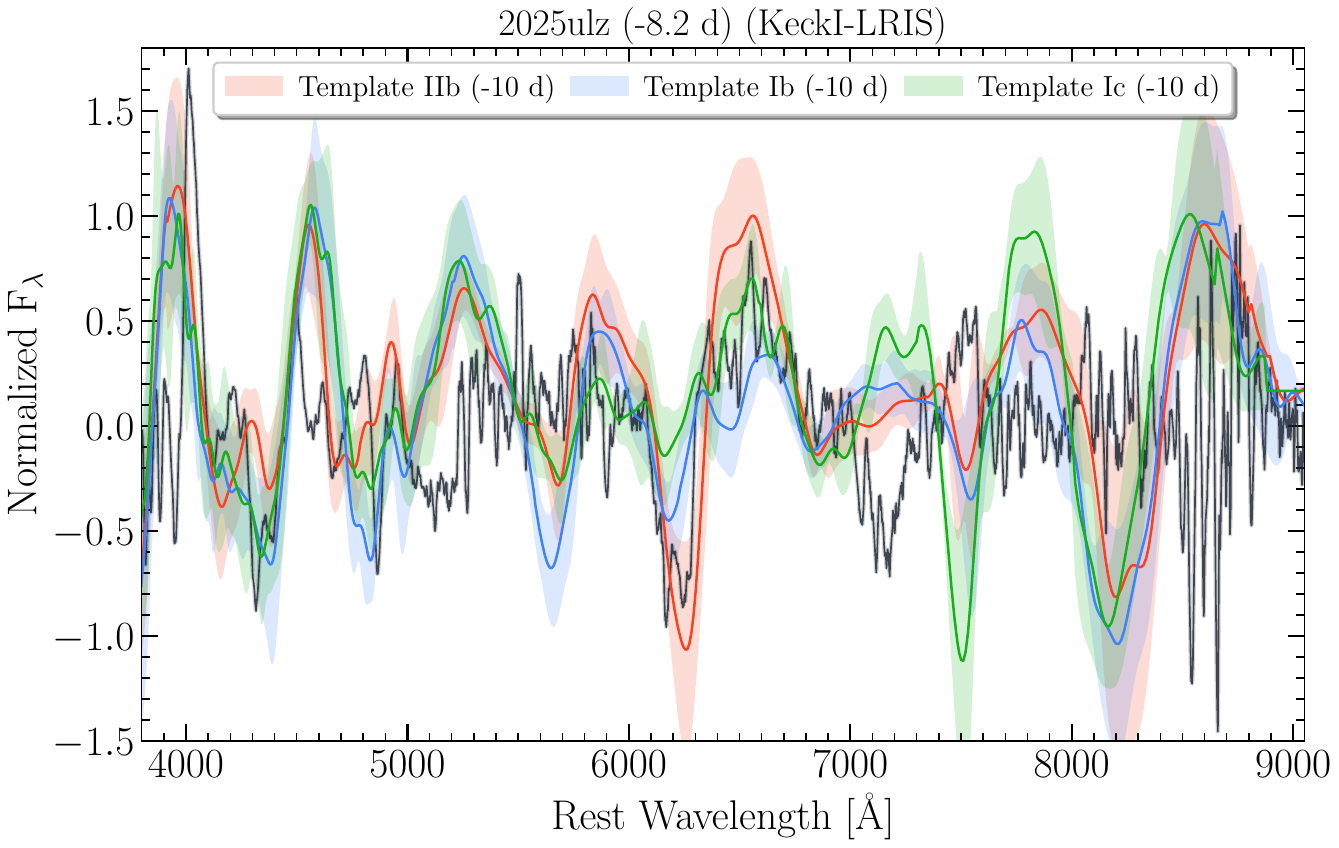} \\
\includegraphics[width=0.5\textwidth]{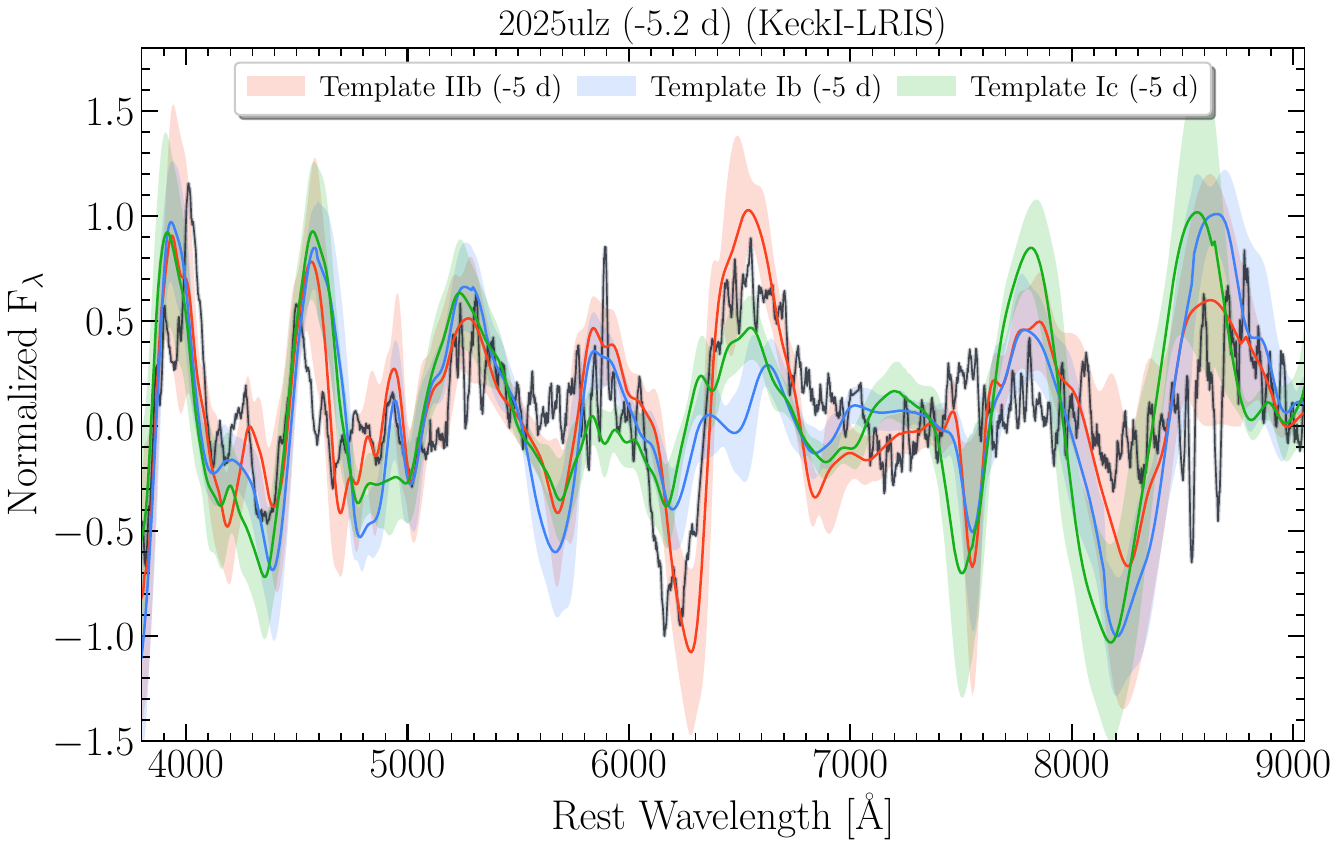}
\includegraphics[width=0.5\textwidth]{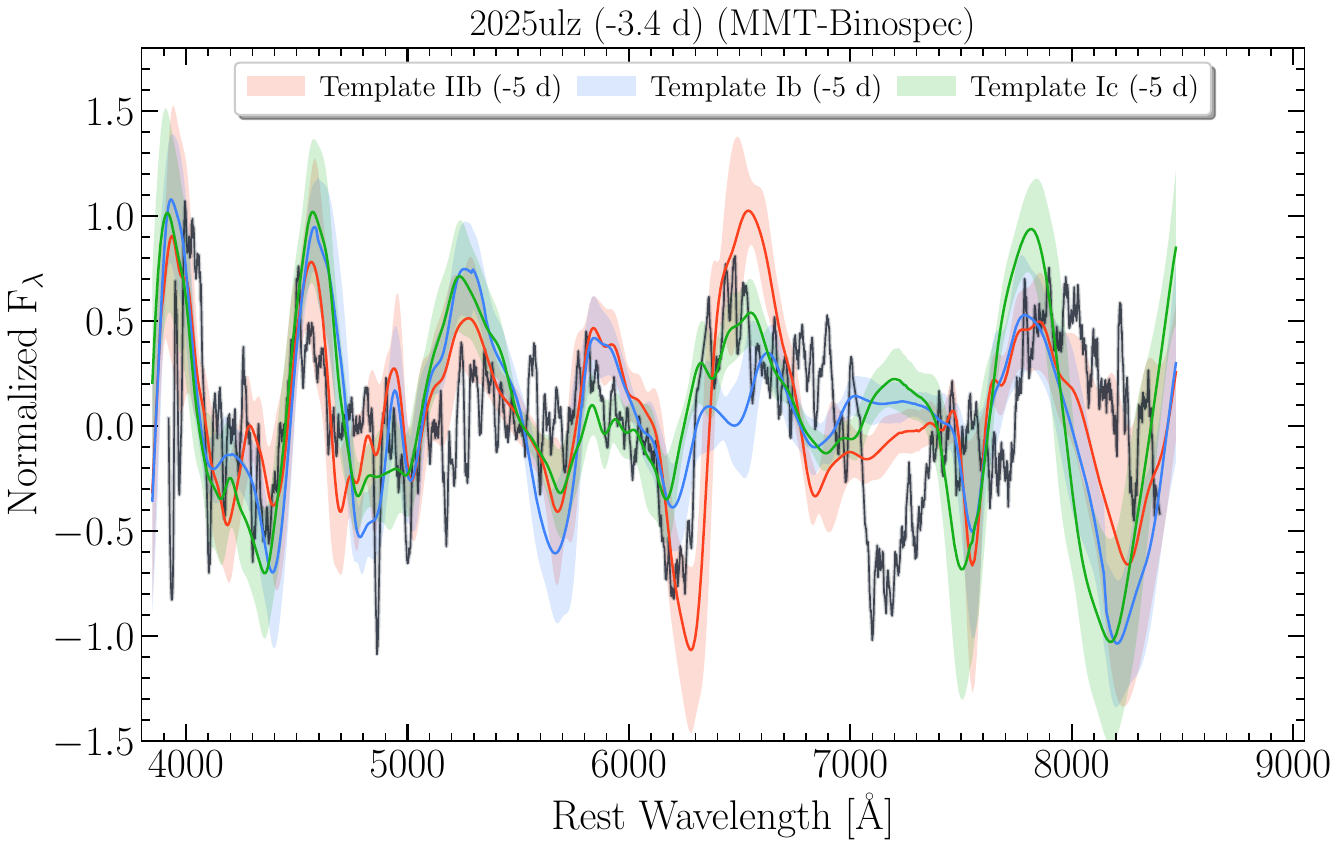}
\includegraphics[width=0.5\textwidth]{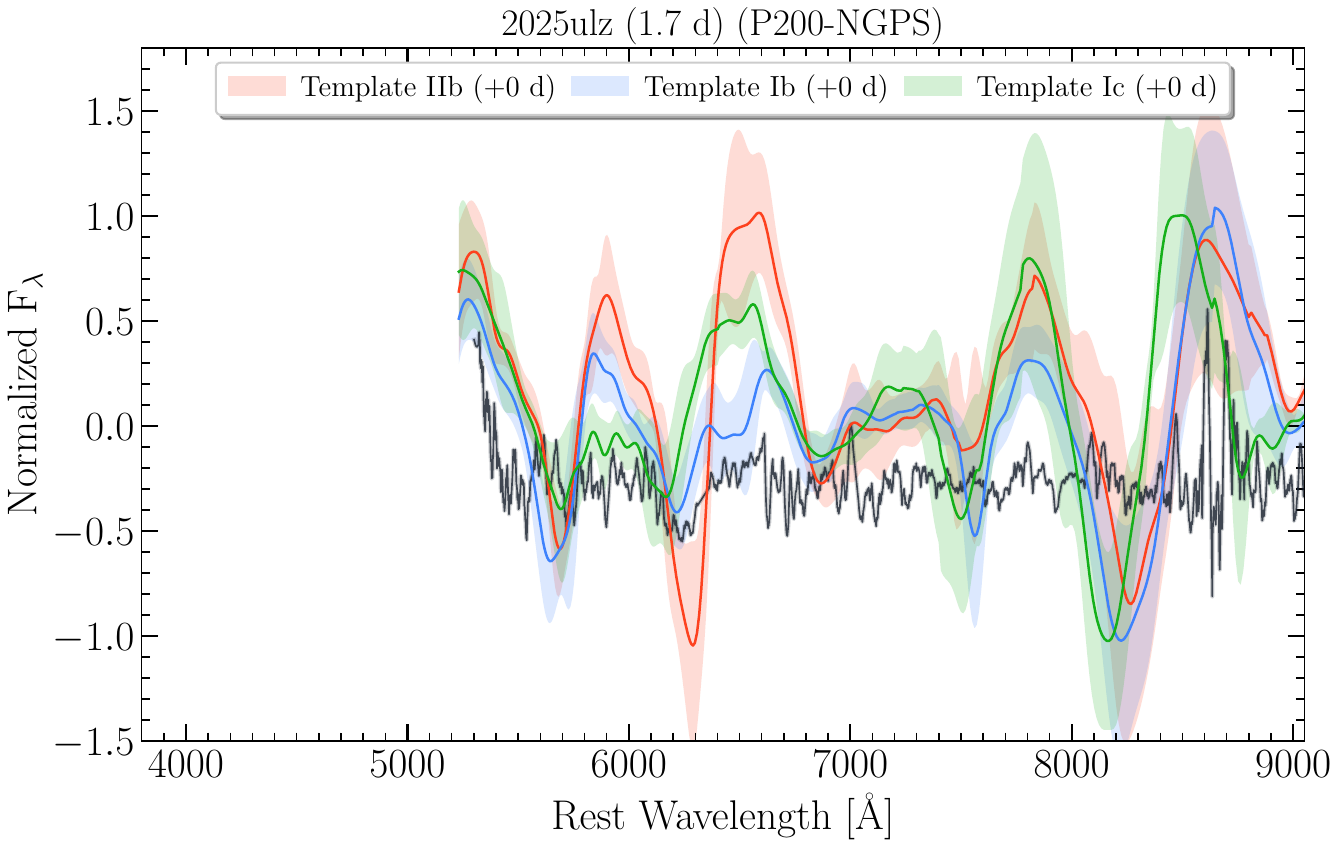}
\includegraphics[width=0.5\textwidth]{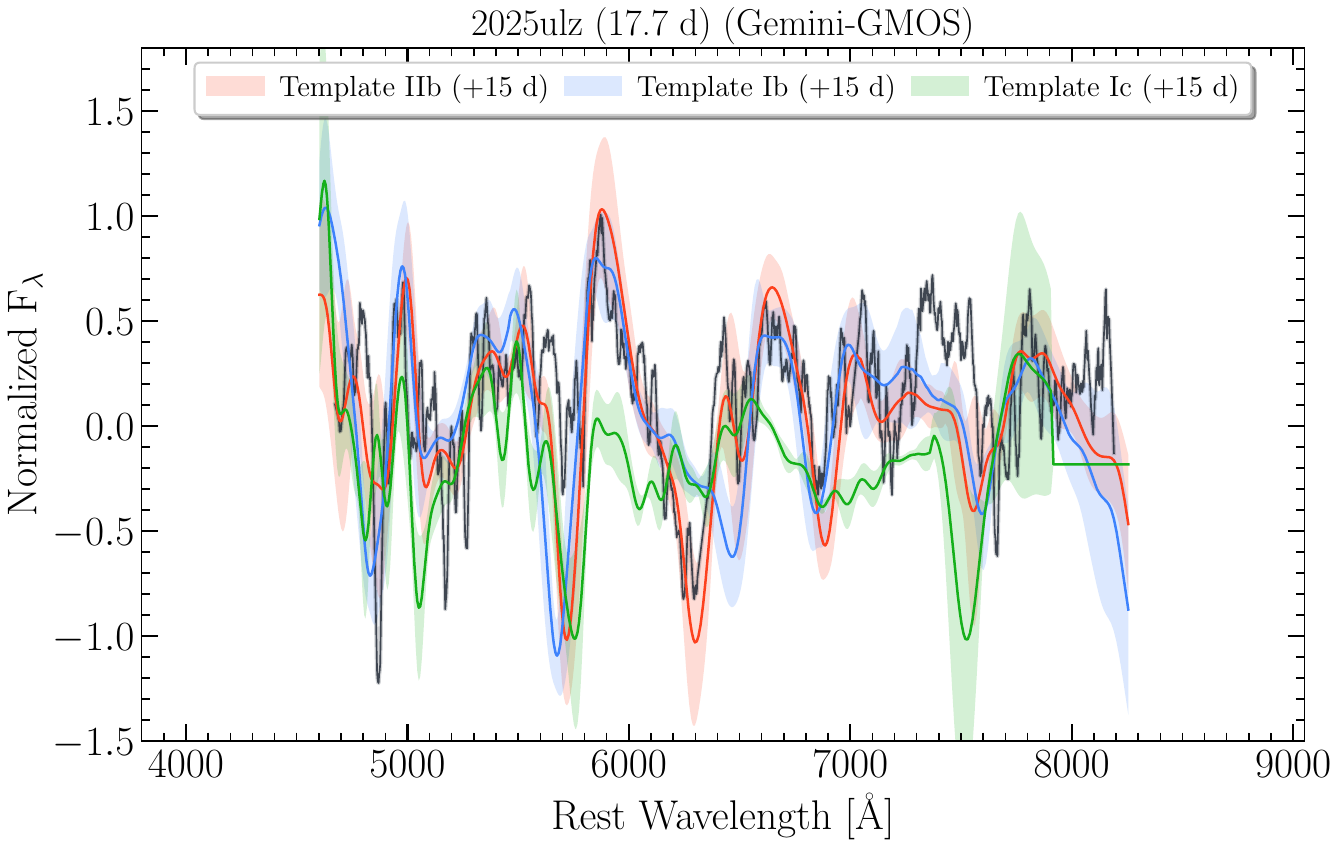}
\caption{Comparing ZTF\,25abjmnps with template spectra of SNe IIb, Ib and Ic from \citet{2016liu, 2016modjaz}. The templates were constructed from phase-binned, flattened mean spectra with the shaded regions representing 1$\sigma$ diversity in each sub-class. Observed spectra were extinction corrected and continuum subtracted after masking the telluric features and narrow host emission lines. The matches to H$\alpha$ before maximum (Panels 2, 3 4) and Helium after maximum (Panel 6) and the absence of O I support the spectroscopic classification as a Type IIb supernova.  
%The lack of He\,I $\lambda$ 5876 suggests that the more prominent feature is H$\alpha$ and not Si\,II and the classification is closer to a IIb than a Ib.
%{\it Figure Credit: Avinash}
}
\label{fig:sntemplates}
\end{figure}

We assume that the early data is dominated by the kilonova with negligible contribution from the afterglow, and thus the kilonova model provides a good description of the early observations. Since the kilonova model fades rapidly after about \( 2 \times 10^5\)~s, we leave a small gap and select data after \( 3\times 10^5\)~s as ``afterglow-dominated'', and perform a fit with \textsc{JETSIMPY} to calculate jet parameters, following the procedures described in \citet{2025arXiv250902769S}. We find a smaller viewing angle and go back and re-do the kilonova fit with additional \textsc{possis} simulations with a finer grid resolution in terms of viewing angle. With this iterative joint fit, we find a best-fit model as shown in Figure~\ref{fig:kilo-grb-lc}, and the best-fit parameters are in the figure caption. We note that this is not the only solution - running joint fits using the NMMA framework \citep{Pang2023} yields fits with lower energy.  

%We note that the viewing angle inferred from this fit is different from the one in the kilonova model -- which we address next.

%Since the kilonova models are created from numerical simulations, it is computationally expensive to perform a full joint fit. Instead, we take the best-fit kilonova model from \S\ref{sec:kilonova_component}, and re-calculate the light curve at various viewing angles. We picked the simulation with a viewing angle of 8\degr\ as the ``correct'' kilonova model. 
%We note that the models for viewing angles within \(\pm1\degr\) are consistent within model uncertainties. We subtract the fluxes predicted by this model from the entire data set, and this time fit the jet model to the complete kilonova-subtracted data set. The final best-fit model is shown in Figure~\ref{fig:kilo-grb-lc}, and the best-fit parameters are in the figure caption.

Comparing to the afterglow modeling in \citealt{OConnor2025}, the best-fit model here explores parameter space that is further off-axis and higher circumstellar density. 
However, while the best-fit model appears to roughly explain all the optical photometry, it is inconsistent with the deeper radio upper limits (Corsi et al. in prep, \citealt{OConnor2025}) and marginally inconsistent with deeper X-ray observations \citep{OConnor2025} that are presented in companion papers.
%However, while the best-fit model appears to roughly explain all the optical photometry, we emphasize that this afterglow model is inconsistent with the deeper radio and X-ray observations (O'Connor et al. accepted, Corsi et al. in prep) that is presented in companion papers. 
Furthermore, the afterglow interpretation for ZTF\,25abjmnps is also inconsistent with the detailed photometric color evolution and the P-Cygni features observed in the optical spectra which we discuss next. Our conclusion that ZTF\,25bjmnps is not a canonical GW170817-like kilonova is consistent with independent analysis in multiple papers \citep{Gillanders2025,Franz2025,Yang2025,Hall2025sn,OConnor2025}. The joint kilonova and afterglow fit result does underline the need to obtain spectroscopic follow-up, radio follow-up and X-ray follow-up for future events and not rely on optical photometry alone. 

%While the best-fit model appears to explain all the optical photometry, we emphasize that this afterglow model is inconsistent with deeper radio (Corsi et al. in prep) and deeper X-ray observations (O'Connor et al. in prep) that will be presented in companion papers. The afterglow interpretation is also inconsistent with the P-Cygni feature observed in the optical spectra which we discuss next. The joint kilonova and afterglow fit result underlines the need to obtain spectroscopic follow-up, radio follow-up and X-ray follow-up for future events and not rely on optical photometry alone. 

%%Vishwajeet and Aditya - add figure here

\begin{figure}[hbt!]
\centering
\includegraphics[width=0.9\textwidth]{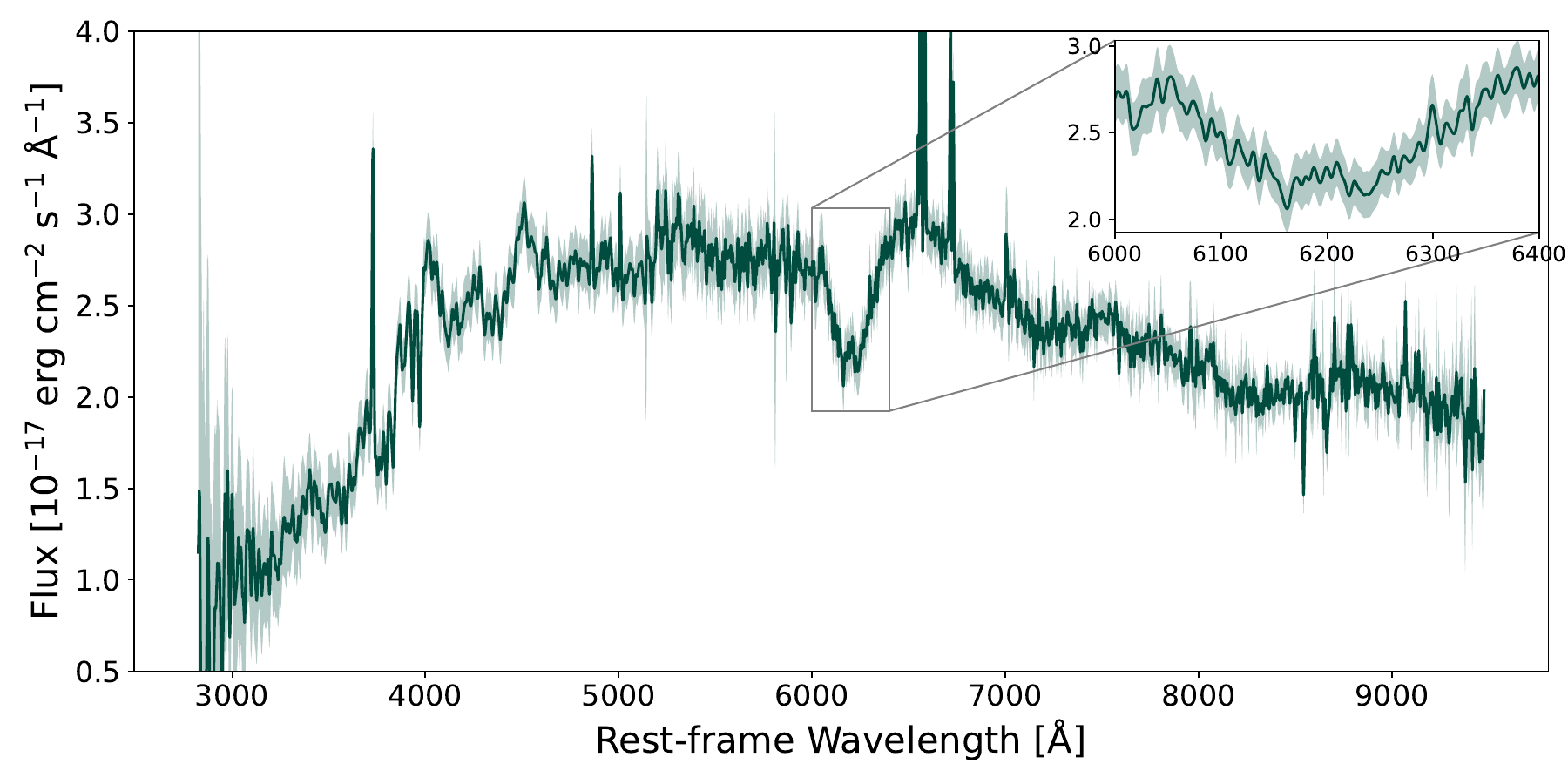}
\caption{Spectrum before subtracting host light. Note the zoom-in on the P-Cygni feature showing the W-shaped profile instead of a U-shaped profile. 
%{\it Figure Credit: Xander}
}
\label{fig:specshape}
\end{figure}

\begin{figure}[hbt!]
\includegraphics[width=0.5\textwidth]{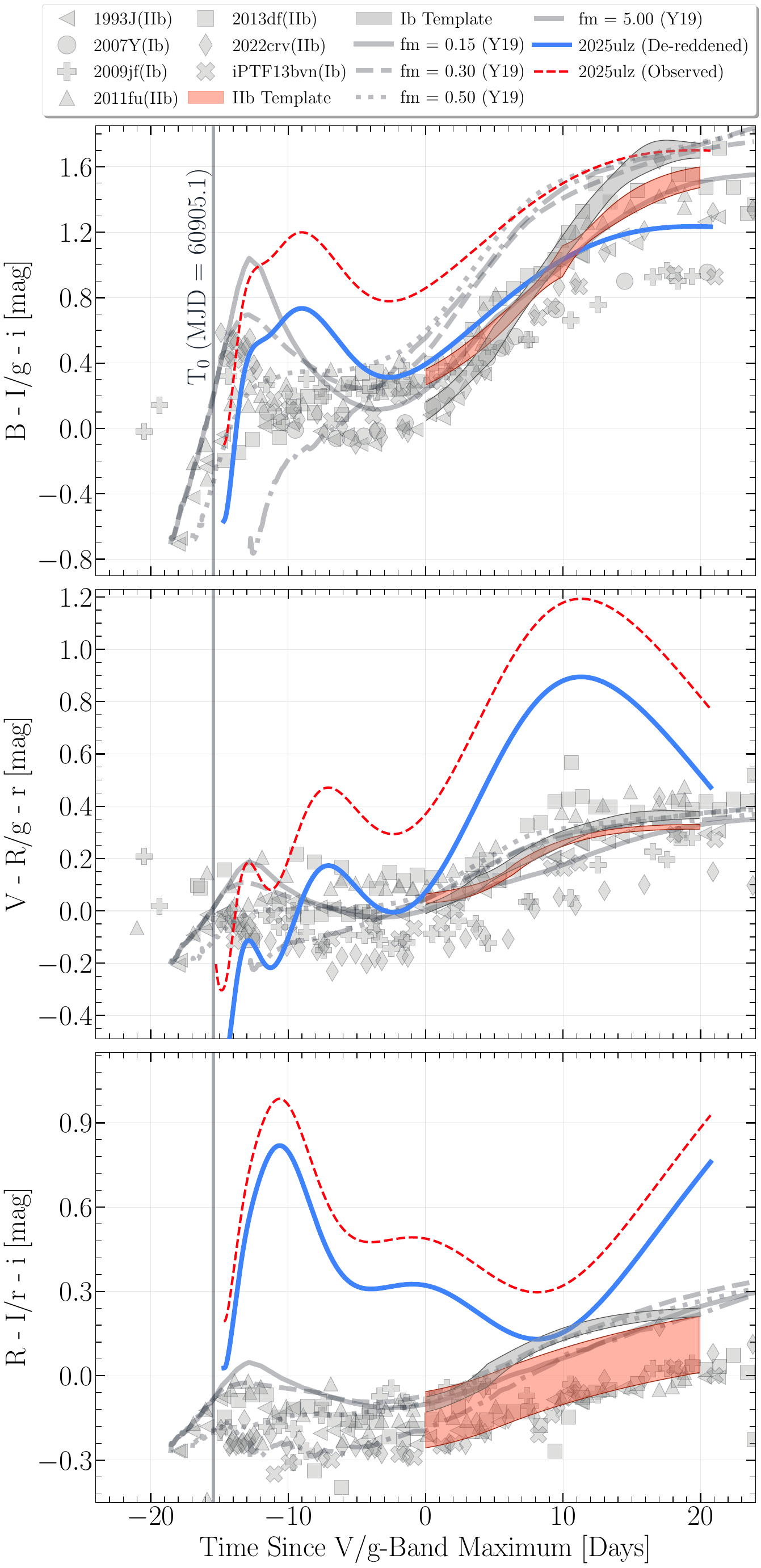} 
\includegraphics[width=0.5\textwidth]{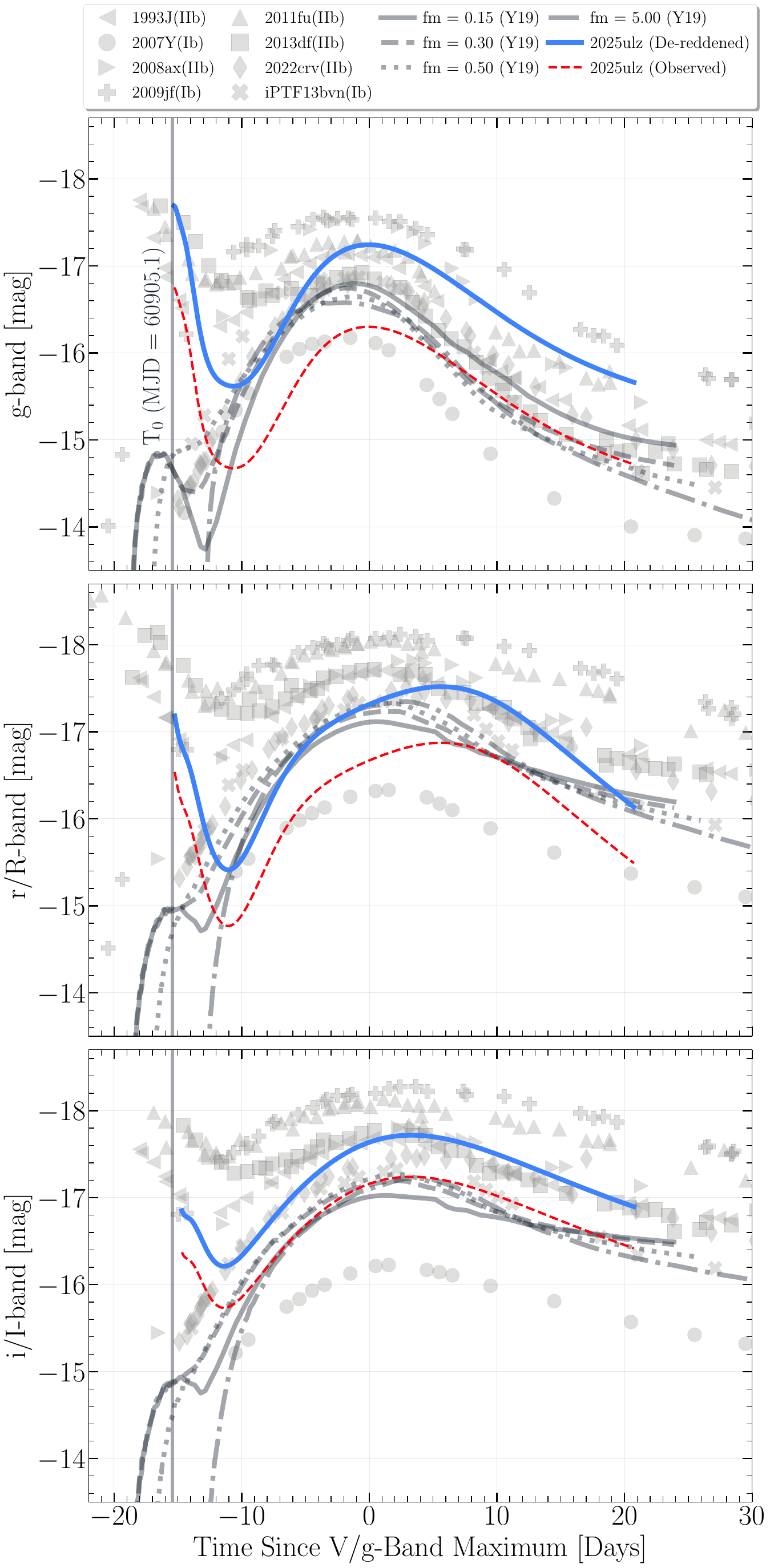}
\caption{{\bf Left Panel:} Comparing the color evolution of ZTF\,25abjmnps with canonical Type IIb supernovae and some templates (observed and dereddened curves are shown as dashed and solid lines). Also shown for comparison are synthetic color curves of SN Ib models by \citet{2019yoon} with varying degrees of $\rm ^{56}Ni$ mixing (higher fm is stronger $\rm ^{56}Ni$ mixing). 
%Results are similar when comparing to the Taddia et al. 2018 sample and the ZTF IIb sample. 
{\bf Right panel:} Comparing the light curve of ZTF\,25abjmnps to the same literature sample of well-studied Type IIb and Type Ib supernovae. Note the steeper decline of the g-band and r-band light curve.
%{\bf Right panel:} Hydrodynamic model fit to ZTF\,25abjmnps using a SN IIb progenitor model with an extended envelope radius $\mathrm{R}_{env}\approx100\mathrm{R}_\odot$, $\mathrm{M}_{ej}\sim2.6\mathrm{M}_\odot$, $\mathrm{M}_{env}\sim0.04\mathrm{M}_\odot$, $\mathrm{M}_{^{56}Ni}\sim0.06\mathrm{M}_\odot$, $\mathrm{E}_{k}=1.0\mathrm{foe}$.
%{\bf Right panel:} Tyler's plot comparing to Christoffer's kilonova impostor sample of faint and fast IIb.
%{\it Figure Credit: Avinash}
}
\label{fig:supernova}
\end{figure}

\section{Is ZTF\,25abjmnps a supernova?}
\label{sec:supernova}

With subsequent follow-up data, there were two major clues suggesting a supernova origin to ZTF\,25abjmnps. First, the appearance and strengthening of a P-Cygni profile in the optical spectra, which if H$\alpha$ at 17,000 km s$^{-1}$ would indicate a core-collapse supernova. Second, a luminous second peak reminiscent of stripped envelope supernovae, specifically Type IIb supernovae, where the first peak is attributed to shock cooling and the second peak to radioactive decay of $^{56}$Ni. 
Our spectroscopic classification below of ZTF\,25abjmnps as a Type IIb supernova is consistent with independent analysis presented in multiple papers \citep{Gillanders2025,Franz2025,Yang2025,Hall2025sn,OConnor2025}. However, a Type IIb classification does not preclude association with S250818k. Thus, next, we compare ZTF\,25abjmnps to a literature sample of well-studied stripped envelope supernovae (Type IIb, Type Ib and Type Ic) to better understand the similarities and differences.

%We compare ZTF\,25abjmnps to two samples of supernovae: a literature sample of well-studied Type IIb supernovae and a ZTF sample of double-peaked Type IIb supernovae. We find several clues that suggest this was an unusual core-collapse supernova, both spectroscopically and photometrically.
Spectroscopically, the first three observations (at +1d, +2d, +3d) show that the continuum gets redder, but do not allow a classification as they are featureless (Figure~\ref{fig:lris}). The next three spectra (at +7d, +9d, +10d) show a strengthening of a P-Cygni profile with an absorption minimum around 6200\,\AA, and the next two spectra (at +16\,d and +31\,d) show a weakening of the same feature (Figure~\ref{fig:lris}). %None of the spectra of ZTF\,25abjmnps (Figure~\ref{fig:spec}) yield a match with rlap$>5$ in the SNID supernova spectral library, perhaps due to host galaxy light contamination. 
We compare the normalized spectra with stripped envelope supernova templates \citep{2016modjaz,2016liu} in Figure~\ref{fig:sntemplates}. The relatively best match is to a Type IIb supernova at $-5$\,d from the second peak with the putative P-Cygni feature being H$\alpha$ at 17,000 km s$^{-1}$ (Figure~\ref{fig:sntemplates}). However, the strength of the same feature is weaker at other phases relative to the templates. We consider whether the putative P-Cygni feature could be He I at 20,000 km s$^{-1}$ as in a Type Ib supernova instead of H$\alpha$, but find that the absence of He I $\lambda$5876 (usually a stronger transition) is inconsistent with this interpretation. We also consider whether the putative P-Cygni feature could be due to Si II at 8,000 km s$^{-1}$ as in a Type Ic supernova. However,  
%and find that the relative strength is a better match at many phases  (Figure~\ref{fig:sntemplates}). 
the absence of the O I $\lambda$7774 line suggests it is not a Type Ic supernova \citep{2019Shivvers}.
The emergence of Helium features, especially He I 5876, in the spectrum after second peak (+17.7\,d) supports the Type IIb classification. We note a confounding detail --- the shape of the putative H$\alpha$ absorption feature was ``W-shape" and not the usual ``U-shape", suggesting a doublet not a singlet line or a more complicated ejecta geometry (Figure~\ref{fig:specshape}). One possibility is that this line is a blend of H$\alpha$ and He I at 20,000 km s$^{-1}$. If so, it is surprising that He I 5876 is not detected at this phase. If this was the Si II doublet, the separation between the absorption dips doesn't match to the same velocity. Overall, ZTF\,25abjmnps is spectroscopically most consistent with being a stripped envelope supernova of Type IIb with some small oddities about line strength and line shape when compared to templates.

Photometrically, we compare the color evolution and light curve evolution to a literature sample of Type IIb and Type Ib supernovae and some theoretical models that vary mixing (Figure~\ref{fig:supernova}). Similar to Type IIb SNe, ZTF\,25abjmnps shows two peaks, albeit with a somewhat steeper evolution after the first peak. However, we find that the $r-i$ color evolution before second peak and the $g-r$ color evolution after second peak is unusual compared to other Type IIb supernovae (Figure~\ref{fig:supernova}, bottom left panel and middle left panel). There are three possibilities for this unusual color evolution. First, we consider the extinction corrections (solid line vs. dashed line in Figure~\ref{fig:supernova}) and find that even using the upper limit on dust from \S~\ref{sec:followup}, we cannot explain the color. Second, we consider whether it could be a k-correction effect, as the literature sample is at significantly lower redshifts than ZTF\,25abjmnps, and the most prominent P-Cygni feature in the spectrum straddles the $r$-and $i$-bands. Quantifying this correction precisely is challenging due to the host galaxy light contamination in the spectra. We roughly estimate that k-correction could explain up to 0.3$\pm$0.1\,mag of $r-i$ color at 5\,days but not the full 0.9\,mag deviation from the literature sample. Third, it is possible that the odd color evolution is due to opacity effects, such as hydrodynamic mixing by disk winds of a small quantity of heavy elements produced by the r-process \citep{Barnes2023} or perhaps even some mixing of Fe-group elements \citep{2019yoon}. A detailed model for the mixing and opacity is outside the scope of this paper. 

Next, to understand the energetics, we fit a radiation hydrodynamics model to the light curve. To create a SN IIb progenitor model, we start with the models developed by \cite{GangLong2022} produced with the Modules for Experiments in Stellar Astrophysics code \citep[\textsc{mesa};][]{Paxton2011, Paxton2013, Paxton2015, Paxton2018, Paxton2019, Jermyn2023}. We use the model with rotation from \citealt{GangLong2022}\footnote{Available at zenodo.org: \cite{GangLong2022_data}}, modified to initial masses between $17~$M$_\mathrm{\odot}$ and $15~$M$_\mathrm{\odot}$, and an orbital period of $300$ days. The evolution of this system results in a SN IIb-like progenitor with an envelope radius $\mathrm{R}_{\rm env}\approx450~\mathrm{R}_\odot$. We use the SuperNova Explosion code \citep[\textsc{snec};][]{Morozova2015} to explore a grid of pre-supernova stellar models based on the binary model described above. This grid covers $\mathrm{E}_\mathrm{kin}=1.0$ to $1.8\times10^{51}$~erg , $\rm M_{{^{56}}Ni}$ 0.08 to 0.18 $\rm M_{\odot}$, $\rm R_{env}$ 100 to 200 $\rm R_{\odot}$, and $\rm M_{ej}$ 3.0 to 3.8 $\rm M_{\odot}$.\footnote{Achieved by modifying the mass excision between 1.2 to 2.0 $\rm{M}_\odot$.} To achieve the range of different envelope radii, we modified the bound envelope radius of the  original $\rm R_{env}$\,=\,450 $\rm R_{\odot}$ model by cutting the model grid appropriately before running \textsc{snec}. We exploded all models in the grid using a thermal bomb with $\rm ^{56}Ni$ mixed all the way through the remaining star after mass excision. We also allow for a shift $\Delta T$ with respect to $T_0$ of S250818k, and a varying E(B-V) during our fitting.

\begin{figure}[hbt!]
\centering
\includegraphics[width=0.8\textwidth]{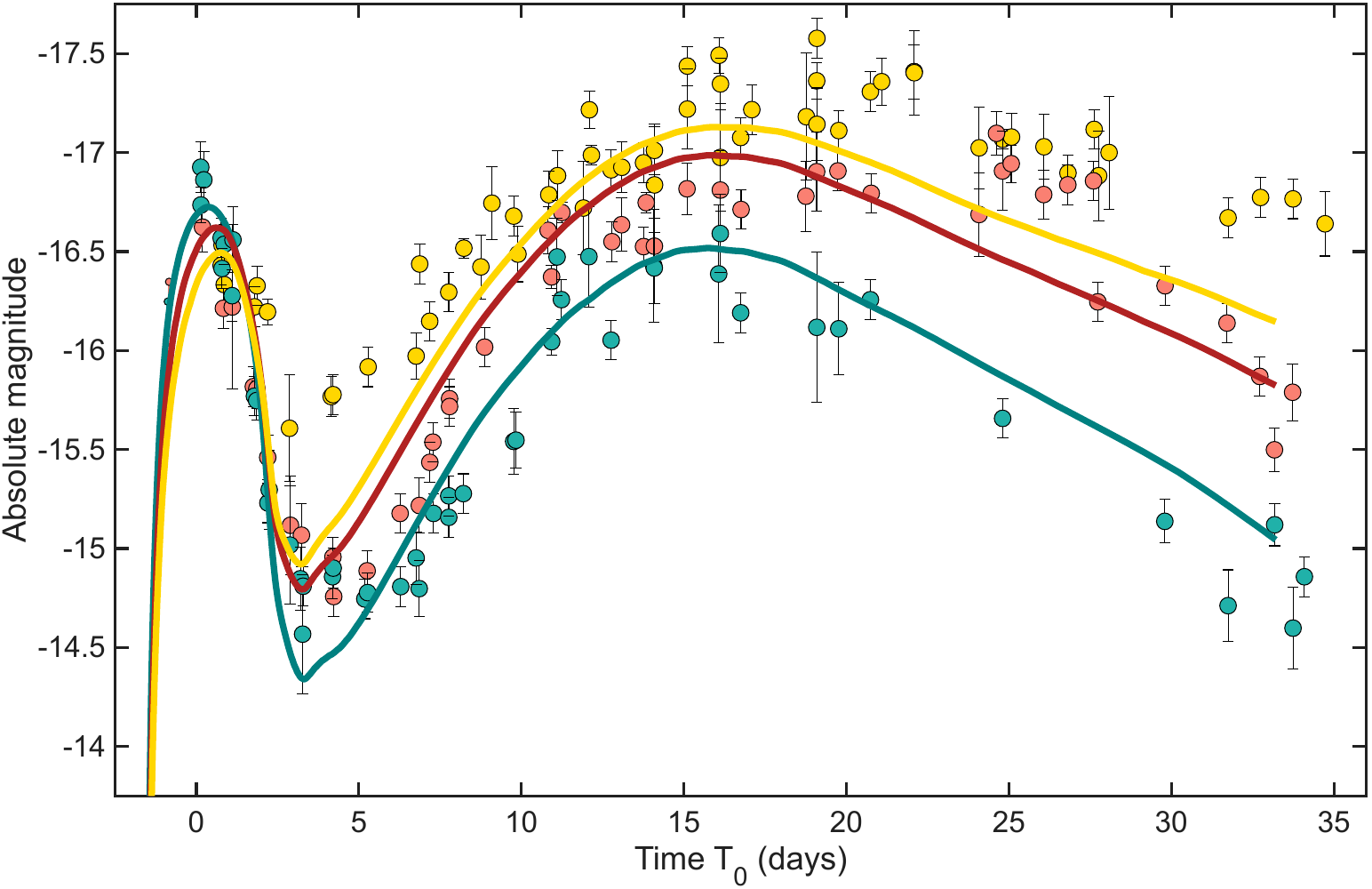}
\caption{Hydrodynamic model fit to ZTF\,25abjmnps using a SN IIb progenitor model with an extended envelope radius $\mathrm{R}_{\rm env}\approx140\mathrm{R}_\odot$, $\mathrm{M}_{\rm ej}\sim3.0\mathrm{M}_\odot$, $\mathrm{M}_{\rm env}\sim0.06\mathrm{M}_\odot$, $\mathrm{M}_{\rm ^{56}Ni}\sim0.1\mathrm{M}_\odot$,
$\mathrm{E}_{k}=1.6\mathrm{foe}$.
$\mathrm{\Delta T}=-1.4\mathrm{days}$.
$\mathrm{E}_{B-V}=0.2\mathrm{mag}$. Note the colors are $g$-band (green), $r$-band (red) and $i$-band (yellow). The pre-explosion upper limit used in the fit is at 3$\sigma$. 
%{\it Figure Credit: Christoffer}
}
\label{fig:hydro}
\end{figure}

The resulting lightcurves were compared to ZTF\,25abjmnps. We find that the optimal\footnote{Based on $\chi^2$ fitting.} explosion parameters for ZTF\,25abjmnps within this grid are $\mathrm{R}_\mathrm{env}\approx140~\mathrm{R}_\odot$, $\mathrm{M}_\mathrm{env}\sim0.06~\mathrm{M}_\odot$, 
$\mathrm{M}_\mathrm{ej}\sim3.0~\mathrm{M}_\odot$, $\mathrm{M}_{^{56}\mathrm{Ni}}\sim0.1~\mathrm{M}_\odot$,
$\mathrm{E}_\mathrm{k}=1.6~\mathrm{foe}$, $\mathrm{\Delta T}=-1.4~\mathrm{days}$ and $\mathrm{E}_{B-V}=0.2~\mathrm{mag}$. This MESA$+$SNEC model is shown in Figure~\ref{fig:hydro}. We cannot measure reliable photospheric expansion velocities from our spectral sequence (e.g., from Fe II), but the best fitting model velocities were checked to not be higher than the velocities measured from the H$\alpha$ absorption minima.

We compare the explosion parameters to those of previously modeled SNe IIb (e.g., \citealt{Taddia2018,Sravan2020}). For example, SN~1993J had $\mathrm{M}_\mathrm{ej}\sim2.5\text{–}3.0~\mathrm{M}_\odot$, $\mathrm{M}_{^{56}\mathrm{Ni}}\sim0.08~\mathrm{M}_\odot$, $\mathrm{E}\mathrm{k}\sim1.0\text{–}1.2$~foe, and a progenitor radius of $\sim600~\mathrm{R}_\odot$ \citep{Nomoto1993,Woosley1994}. SN~2011dh was modeled with $\mathrm{M}_\mathrm{ej}\sim1.8\text{–}2.5~\mathrm{M}_\odot$, $\mathrm{M}_{^{56}\mathrm{Ni}}\sim0.06~\mathrm{M}_\odot$, $\mathrm{E}\mathrm{k}\sim0.6\text{–}1.0$~foe, and a progenitor radius of $\sim200~\mathrm{R}_\odot$ \citep{Bersten2012,Ergon2014}. Similarly, SN~2013df showed $\mathrm{M}_\mathrm{ej}\sim2.0\text{–}3.5~\mathrm{M}_\odot$, $\mathrm{M}_{^{56}\mathrm{Ni}}\sim0.1~\mathrm{M}_\odot$, $\mathrm{E}\mathrm{k}\sim0.8\text{–}1.2$ foe, and a progenitor radius of $\sim550~\mathrm{R}_\odot$ \citep{MoralesGaroffolo2014}. Thus, ZTF\,25abjmnps falls within the canonical Type  IIb SN parameter space, except for an apparently smaller envelope radius and somewhat higher kinetic energy, though there may be a selection bias against small radii since their duration is fast and rarely well enough sampled for detailed modeling. Based on the derived parameters, ZTF\,25abjmnps is more likely to have a binary progenitor.

%\begin{figure}[hbt!]
%\includegraphics[width=0.5\textwidth]{PLOT_2025ulz_CompareColorswithNormalSESNe.pdf} 
%\includegraphics[width=0.5\textwidth]{PLOT_2025ulz_ColorComparisonCanonicalIIb_B-I.pdf}
%\caption{{\bf Left Panel}: Comparison of the V-R/g-r and R-I/r-i color curves comparison of normal SESNe from literature with ZTF\,25abjmnps. The dereddened color curves are shown in solid lines and the observed colors are shown with dashed lines.
%{\bf Right Panel}: The r-i color evolution of ZTF\,25abjmnps compared with the synthetic R-I color curves of SN Ib models by \citet{2019yoon}. The line curves indicate varying degrees of $\rm ^{56}Ni$ mixing, fm as shown in the legend. An increasing value of fm implies a stronger $\rm ^{56}Ni$ mixing. {\it Figure Credit: Avinash}
%}
%\label{fig:sn2}
%\end{figure}

Finally, to estimate the chance coincidence of discovering a young Type IIb supernova in the observed snapshot gravitational wave volume within one day of explosion, we do the following back-of-the-envelope calculation: $1-$exp$(-$Volume*Age*Rate$)$. To estimate the Volume, we consider a spherical shell between 150\,Mpc and 400\,Mpc (corresponding to the GW constraints) and a fractional all-sky volume of 0.008962 (corresponding to the 369.7 sq deg searched) and get $2.276\times\,10^{6}$ Mpc$^{3}$. For Age, we assume the transient is within one day of explosion, so a time window of 1/365 yr. For the Rate of Type IIb, there is an unresolved discrepancy between magnitude-limited survey estimates and volume-limited survey estimates. We assume the Type IIb fraction of 12\% relative to Type II from \citealt{Li2011} and the Type II rate from \citealt{Das2025} of $3.9\times10^{-5}\,$Mpc$^{-3}\,$yr$^{-1}$. Thus, our net chance coincidence estimate is crudely 2.9\%. 
%%10.6% of 6.2e-5, Smith et al. 2011 LOSS fraction of IIb 
A more recent estimate of the Type IIb volumetric rate by ASAS-SN was $0.8^{+0.6}_{-0.4}\times10^{-5}\,$Mpc$^{-3}\,$yr$^{-1}$ \citep{Pessi2025} -- this gives a chance coincidence rate of 4.9$^{+3.5}_{-2.5}$\%.  
We caution that this is not a robust estimate for many reasons; we have not folded in the joint significance of the EM and GW, trials factors, the detection efficiency of ZTF, or the fraction of IIb that show the peculiar properties of ZTF\,25abjmnps. While a more detailed estimate is beyond the scope of this paper as it awaits the offline GW analysis, this back-of-the-envelope estimate tells us that the chance coincidence of finding a young stripped envelope supernova in the gravitational wave localization is not negligible. As described in Appendix B, we further re-analyze our candidate vetting procedure to thoroughly check if there may be any other young stripped envelope supernovae that may be coincident with S250818k and ZTF\,25abjmnps remains the most compelling candidate.

\section{Discussion: are ZTF\,25abjmnps and S250818k related by a superkilonova model?}
\label{sec:discussion}

%We conclude that ZTF\,25abjmnps cannot be trivially classified as either a supernova or a kilonova. 

To summarize, ZTF\,25abjmnps is an optical transient that is spatially and temporally coincident with the gravitational wave signal. While the optical light curve fits a kilonova and afterglow model, the optical spectral features and radio/X-ray upper limits are inconsistent with this picture. While the optical spectra and light curves are most similar to those of  a Type IIb supernova, there are some differences in color evolution and spectral feature strength/shape when compared to the literature sample of Type IIb supernovae. The key gravitational wave clue is that this may represent the merger of neutron stars lighter than a solar mass.  Next, we explore a potential tantalizing multi-messenger association in the context of a superkilonova model.
 
In principle, stable neutron stars can exist and be as small as 0.1 $\mathrm{M}_\odot$ (e.g., \citealt{Haensel2002}). However, the formation of a sub-solar neutron star presents a major challenge to stellar evolution. Detailed modern simulations of the core collapse and supernova explosion of (slowly spinning) massive stars predict a robust lower limit to the neutron star mass of 1.2 $\mathrm{M}_\odot$, similar to the Chandrasekhar mass of the progenitor’s iron core \citep{Muller2025}. A similar lower limit applies to the masses of neutron stars formed from the accretion-induced collapse of a white dwarf (e.g., \citealt{NomotoKondo1991}). The only solution to forming a sub-solar mass neutron star is to identify a neutron-rich environment as the Chandrasekhar mass is proportional to square of the electron fraction. One such gravitating, neutron-rich environment is possible in the immediate aftermath of the collapse of a rapidly spinning star. It has been proposed that collapsing stars may undergo fission of the collapsing core into two neutron stars instead of one \citep{Durisen1985, Imshennik1998, Davies2002, Postnov2016}, although this is yet to be demonstrated with detailed numerical simulations. Separately, it has also been proposed that if the stellar envelope has sufficient angular momentum, then rather than accreting directly onto the central compact object, it would initially form a centrifugally-supported disk of material. 

A sufficiently massive and large disk can become gravitationally unstable \citep{Toomre1964} and undergo fragmentation and collapse to form neutron stars, as a result of runaway neutrino cooling or alpha-particle dissociation \citep{PiroPfahl2007, Metzger2024,Lerner25}. This process is qualitatively similar to the proposed mechanism for forming planets in protoplanetary disks \citep{Gammie2001} or stars in the disks of active galactic nuclei \citep{GoodmanTan2004}. Furthermore, if the disk becomes neutron-rich as a result of electron captures on protons, then the lower Chandrasekhar mass limit could enable the formation of sub-solar mass neutron stars \citep{Metzger2024}. This was recently illustrated explicitly using shearing-box hydrodynamic simulations by \citet{ChenMetzger2025}, who found that fragmentation into a spectrum of objects of mass $\approx$0.01--1\,$\mathrm{M}_\odot$ is achieved for disks with high accretion rates. If the disk fragments into multiple neutron stars, these bodies may become paired into tight binaries, either as a result of fissioning from a single collapsing clump (e.g., \citealt{Nesvorny2010}) or through gas drag friction (e.g., \citealt{DodiciTremaine2024}). Subsequent coalescence of these binaries, possibly after a delay of minutes to hours after the collapse, offer a potential source of gravitational wave emission in near coincidence with the supernova \citep{Metzger2024}. Alternatively, if the disk fragments into a single sub-solar neutron star, it will merge with the central (ordinary mass) neutron star or low-mass black hole created by the core collapse, creating a single gravitational wave signal.

We posit that this idea of fragmentation in a collapsing core's accretion disk can qualitatively explain a possible multi-messenger association between S250818k and ZTF\,25abjmnps. First, the low chirp mass of S250818k leaves open the possibility that we are seeing the merger of one or two sub-solar neutron stars, either with themselves or with the central compact object left over from the explosion. More details on the masses of the two components and revised false alarm rate with an offline analysis of the gravitational wave strain will need to await future publication by the LIGO-Virgo-KAGRA collaboration. Second, the similarity of ZTF\,25abjmnps to stripped envelope supernovae suggests that the progenitor star might have interacted with a binary, leaving it with enough angular momentum to form such an accretion disk that can fragment upon collapse. Although past work has focused on this possibility in the context of collapsars, there is no fundamental reason that rapid rotation and associated disk formation is limited to the collapse of completely stripped stars or broad-line Type Ic supernovae in particular. Indeed, the mechanisms for spinning up the cores of massive stars at late stages of evolution may be diverse (e.g., \citealt{Cantiello2007,Tsuna2025}) and in some cases may occur for progenitors with moderate amounts of hydrogen and helium. For example, a stellar merger that occurs soon before core-collapse could both spin up the core and give rise to a Type IIb-like supernova (e.g., \citealt{Yoon2017, Lohev2019}). Third, the unusual color evolution may be indicative that a small amount of r-process elements get mixed and contribute to the opacity for some time until the Nickel dominates \citep{Kasen2013,Barnes2023}. Fourth, the actively star-forming local environment of the host galaxy of ZTF\,25abjmnps is a suitable location for such core-collapse events.  

Establishing a firmer association between S250818k and ZTF\,25abjmnps requires more detailed theoretical modeling and sensitive late-time observations. Late-time monitoring of the light curve could be compared to what is expected from heating by Nickel-56. Nebular spectroscopy in the infrared (especially with the James Webb Space Telescope), would directly constrain ejecta composition. Perhaps the high density of the surrounding medium in actively star-forming regions could aid radio detections at late-time. Additional radio and X-ray data would constrain whether or not there is a late-time relativistic non-thermal component. If the accretion disk or disk-embedded merger events power relativistic jets, this could impart asymmetry to the supernova explosion along the rotation axis, and create a non-thermal afterglow if the jet can break through the envelope of the star. On the other hand, a larger envelope could stifle any accretion-powered jet, rendering signatures of the central engine less apparent than in traditional GRBs.  While late-time radio/X-ray detections would support this idea, late-time radio/X-ray non-detections would not rule out this superkilonova model. Perhaps the most vivid proof of such a superkilonova model would be if the gravitational wave signature itself showed signs of multiple mergers, for example, a sub-solar neutron star merger followed by a neutron star black hole merger. A neutron star black hole merger is only expected in the superkilonova picture if the accretion disk fragmented to form the sub-solar neutron stars. It is not expected in the superkilonova picture if the collapsing core directly fissioned into two neutron stars which merged or if the fragmentation formed only one sub-solar neutron star that merged with a central neutron star.

%We qualitatively explore the idea proposed by Metzger that a core-collaps supernova with sufficient rotational kinetic energy formed an accretion disk during collapse that fragmented. These fragmented disks led to the formation of two sub-solar neutron stars and their subsequent merger is what was seen by the GW interferometers. All the data in-hand appears to be qualitatively consistent with this idea. 

Looking back to previous multi-messenger searches of neutron star mergers, stripped envelope supernovae have been rejected as associated at least twice. First, DG19wxnjc (AT2019npv)  was a Type Ib-like supernova that was spatially and temporally coincident with the neutron star black hole merger GW190814 \citep{Andreoni2020}. However, the age of the supernova was unconstrained in this case and the gravitational event could have been a binary black hole merger. Second, AT2019wxt was an ultra-stripped supernova that was rejected during follow-up of S191213g \citep{Shivkumar2023,Agudo2023}. However, the astrophysical significance of the gravitational wave data was deemed questionable. More generally, a re-examination of past sub-threshold events may be warranted in the context of the superkilonova model. 

Looking ahead, there are many ways to test whether or not this potential multi-messenger association between S250818k and ZTF\,25abjmnps is correct. With future more sensitive GW interferometers in O5 and beyond, there should be many more detections of sub-solar binary neutron star mergers at higher significance and better localization. Looking for an associated stripped-envelope core-collapse supernova-like signature would be straightforward even to further distances with the next generation of surveyors: e.g., Vera C. Rubin Observatory \citep{rubin,Andreoni2022}, the Roman Space Telescope \citep{roman}, DSA-2000 \citep{dsa2000, Dobie2021}, the UVEX satellite \citep{uvex,Criswell2025}, Cryoscope in Antarctica \citep{cryoscope}. Any transient that looks like a young, stripped-envelope supernova coincident with a sub-solar neutron star merger or a neutron star black hole merger should be followed up extensively at all wavelengths. More detailed theoretical modeling of a superkilonova, especially light curve and spectra predictions, would help the observers optimize their follow-up observations. Continuing to have information such as the binned chirp mass and HasSSM in low-latency is an important step towards multi-messenger co-operation and would sharpen the telescope response. We remind the reader that all the optical photometry of ZTF\,25abjmnps appears consistent with a canonical kilonova and afterglow model --- spectroscopy, infrared, radio and X-ray data are essential to rule out this scenario. We caution against a rushed conclusion and encourage collecting and analyzing the full panchromatic dataset necessary to more firmly establish a multi-messenger association.  

In summary, we encourage the community to approach the future of multi-messenger astrophysics with a wide-open view to new possibilities. When GW170817 happened in our backyard, the evidence for the multi-messenger association was vivid as the distinction between a kilonova and a supernova was vast. Future multi-messenger events may not look like GW170817, will likely be much further away than GW170817 and may even show similarities to supernovae. Nevertheless, there is an opportunity here to discern a ``veritable multi-messenger symphony" \citep{Metzger2024} of a superkilonova --- a core-collapse supernova, a neutron star merger, and even a neutron star black hole merger at the same time. 

%\section{Other companion paper possibilities}
%\begin{itemize}
%%\item Ruling out the other 57 candidates in the localization?
%%\item How unusual and extreme is this as a supernova? Rates using simsurvey and properties using ZTF core-collapse sample
%%\item Kilonova NMMA modeling?
%\item Host galaxy properties - Xander
%\item Radio (longer time-scale) - Alessandra
%\item X-rays (longer time-scale) - Brendan
%\end{itemize}

%\section{Figures}
%\begin{enumerate}
%\item Discovery Image as zoom in to LVK localization (Use P48 data or LRIS data or Hubble data) - eye candy - Robert?
%\item Light Curve Figure (similar to Mattia figure with GW170817 plotted?)
%\item Light Curve Table (in overleaf format) 
%\item LRIS+NIRES Spectroscopy Collage + GW170817 + models - Xander and Viraj?
%\item Bolometric light curve, blackbody fits and SED evolution
%\item Model fits to light curve - Mattia?
%\item NMMA figures - Michael and Shreya
%\end{enumerate}

\section{Acknowledgements}

The authors thank the anonymous referee for their thoughtful feedback. MMK thanks E. Nakar and K. Hotokezaka for valuable real-time brainstorming discussions. Mahalo to the W. M. Keck Observatory staff, especially M. Lundquist and J. O'Meara, and many observers for their full co-operation with multiple Target Of Opportunity interruptions. MMK thanks the participants of the ZTF Theory Network meeting at Oak Creek, especially Sterl Phinney, Eliot Quataert, Lars Bildsten, Daniel Brethauer, Wynn Jacobson-Galan and Ryan Chornock, for many stimulating discussions and this research benefited from interactions supported by the Gordon and Betty Moore Foundation through Grant GBMF5076. 

A. Singh acknowledges support from the Knut and Alice Wallenberg Foundation through the ``Gravity Meets Light" project. B.~D.~Metzger acknowledges partial support from the National Science Foundation (grant AST-2406637) and the Simons Foundation (grant 727700). The Flatiron Institute is supported by the Simons Foundation. DK is supported in part by the U.S. Department of Energy, Office of Science, Office of Nuclear Physics, DE-AC02-05CH11231, DE-SC0004658, and DE-SC0024388, and by a grant from the Simons Foundation (622817DK). B.O. is supported by the McWilliams Postdoctoral Fellowship in the McWilliams Center for Cosmology and Astrophysics at Carnegie Mellon University. AP, TC, LH are supported by NSF Grant No. 2308193. M. Bulla acknowledges the Department of Physics and Earth Science of the University of Ferrara for the financial support through the FIRD 2024 grant

Based on observations obtained with the Samuel Oschin Telescope 48-inch and the 60-inch Telescope at the Palomar Observatory as part of the Zwicky Transient Facility project. ZTF is supported by the National Science Foundation under Grants No. AST-1440341, AST-2034437, and currently Award \#2407588. ZTF receives additional funding from the ZTF partnership. Current members include Caltech, USA; Caltech/IPAC, USA; University of Maryland, USA; University of California, Berkeley, USA; University of Wisconsin at Milwaukee, USA; Cornell University, USA; Drexel University, USA; University of North Carolina at Chapel Hill, USA; Institute of Science and Technology, Austria; National Central University, Taiwan, and OKC, University of Stockholm, Sweden. Operations are conducted by Caltech's Optical Observatory (COO), Caltech/IPAC, and the University of Washington at Seattle, USA. SED Machine is based upon work supported by the National Science Foundation under Grant No. 1106171. The ZTF forced-photometry service was funded under the Heising-Simons Foundation grant \#12540303 (PI: Graham). The Gordon and Betty Moore Foundation, through both the Data-Driven Investigator Program and a dedicated grant, provided critical funding for SkyPortal.  

Some of the data presented herein were obtained at Keck Observatory, which is a private 501(c)3 non-profit organization operated as a scientific partnership among the California Institute of Technology, the University of California, and the National Aeronautics and Space Administration. The Observatory was made possible by the generous financial support of the W. M. Keck Foundation. Some of the data presented herein were obtained at Keck Observatory, which is a private 501(c)3 non-profit organization operated as a scientific partnership among the California Institute of Technology, the University of California, and the National Aeronautics and Space Administration. The Observatory was made possible by the generous financial support of the W. M. Keck Foundation. 

The 2m Himalayan Chandra telescope (HCT) located at the Indian Optical Observatory (IAO) at Hanle. We thank the staff of IAO, Hanle and CREST, Hosakote, that made these observations possible. HCT observations were carried out under the ToO program of proposal number HCT-2025-C3-P43. The facilities at IAO and CREST are operated by the Indian Institute of Astrophysics, Bangalore.

This article includes observations made in the Two-meter Twin Telescope (TTT) sited at the Teide Observatory of the Instituto de Astrofísica de Canarias (IAC), that Light Bridges operates on the island of Tenerife, Canary Islands (Spain). The Observing Time Rights (DTO) used for this research were provided by Light Bridges, SL.

Based on observations at Cerro Tololo Inter-American Observatory, NSF’s NOIRLab (NOIRLab Prop. ID 2025A-729671; PI: Palmese), which is managed by the Association of Universities for Research in Astronomy (AURA) under a cooperative agreement with the National Science Foundation. Based on observations obtained at the international Gemini Observatory, a program of NSF's OIR Lab, which is managed by the Association of Universities for Research in Astronomy (AURA) under a cooperative agreement with the National Science Foundation on behalf of the Gemini Observatory partnership: the National Science Foundation (United States), National Research Council (Canada), Agencia Nacional de Investigaci\'{o}n y Desarrollo (Chile), Ministerio de Ciencia, Tecnolog\'{i}a e Innovaci\'{o}n (Argentina), Minist\'{e}rio da Ci\^{e}ncia, Tecnologia, Inova\c{c}\~{o}es e Comunica\c{c}\~{o}es (Brazil), and Korea Astronomy and Space Science Institute (Republic of Korea). The data were acquired through the Gemini Observatory Archive at NSF NOIRLab and processed using DRAGONS (Data Reduction for Astronomy from Gemini Observatory North and South). The authors wish to recognize and acknowledge the very significant cultural role and reverence that the summit of Maunakea has always had within the indigenous Hawaiian community.

This paper contains data obtained at the Wendelstein Observatory of the Ludwig-Maximilians University Munich. We thank Christoph Ries, Michael Schmidt and Silona Wilke for performing the observations. Funded in part by the Deutsche Forschungsgemeinschaft (DFG, German Research Foundation) under Germany's Excellence Strategy – EXC-2094 – 390783311. 

MMT Observatory access was supported by Northwestern University and the Center for Interdisciplinary Exploration and Research in Astrophysics (CIERA). C.L. and~A.A.M.~are supported by DoE award \#\,DE-SC0025599, while A.A.M.~is also supported by Cottrell Scholar Award \#\,CS-CSA-2025-059 from Research Corporation for Science Advancement. N.R.~is supported by NSF award \#\,2421845.

The Liverpool Telescope is operated on the island of La Palma by Liverpool John Moores University in the Spanish Observatorio del Roque de los Muchachos of the Instituto de Astrofisica de Canarias with financial support from the UK Science and Technology Facilities Council.

This research used data obtained with the Dark Energy Spectroscopic Instrument (DESI). DESI construction and operations is managed by the Lawrence Berkeley National Laboratory. This material is based upon work supported by the U.S. Department of Energy, Office of Science, Office of High-Energy Physics, under Contract No. DE–AC02–05CH11231, and by the National Energy Research Scientific Computing Center, a DOE Office of Science User Facility under the same contract. Additional support for DESI was provided by the U.S. National Science Foundation (NSF), Division of Astronomical Sciences under Contract No. AST-0950945 to the NSF’s National Optical-Infrared Astronomy Research Laboratory; the Science and Technology Facilities Council of the United Kingdom; the Gordon and Betty Moore Foundation; the Heising-Simons Foundation; the French Alternative Energies and Atomic Energy Commission (CEA); the National Council of Humanities, Science and Technology of Mexico (CONAHCYT); the Ministry of Science and Innovation of Spain (MICINN), and by the DESI Member Institutions: \url{www.desi.lbl.gov/collaborating-institutions}. The DESI collaboration is honored to be permitted to conduct scientific research on I’oligam Du’ag (Kitt Peak), a mountain with particular significance to the Tohono O’odham Nation. Any opinions, findings, and conclusions or recommendations expressed in this material are those of the author(s) and do not necessarily reflect the views of the U.S. National Science Foundation, the U.S. Department of Energy, or any of the listed funding agencies.

\bibliography{references}{}
\bibliographystyle{aasjournalv7}

\appendix
\section{More information about Follow-up Telescopes}

\paragraph{Fraunhofer Telescope at Wendelstein Observatory (FTW)} We observed with the Three Channel Imager (3KK; \citealt{lang2016wendelstein}) instrument mounted on the FTW \citep{2014SPIE.9145E..2DH} in the $g'$, $r'$, $i'$, $z'$, and $J$ bands. The optical CCD and NIR CMOS data were reduced using a custom pipeline developed at Wendelstein observatory \citep{2002A&A...381.1095G, 2025arXiv250314588B}. For the astrometric calibration of the images, we used the Gaia EDR3 catalog \citep{Gaia2021, 2021A&A...649A...2L, gaiaEDR3}. We used the Pan-STARRS1 catalog \citep[PS1;][]{2010SPIE.7733E..0EK} for the optical photometric calibration and the 2MASS catalog \citep{Skrutskie2006} for the $J$ band. Tools from the AstrOmatic software suite \citep{1996A&AS..117..393B, 2006ASPC..351..112B, 2002ASPC..281..228B} were used for the coaddition of each epoch's individual exposures. We used the Saccadic Fast Fourier Transform (SFFT; \citealt{hu_image_2022}) algorithm for image subtraction. 
For subtraction templates we used Legacy Survey DR10 images \citep{2019AJ....157..168D} for the $g$, $r$, $z$ band and PS1 imaging for the $i$ band \citep{2010SPIE.7733E..0EK}. 
FTW was able to provide a quick follow-up within hours of discovery, with two epochs showing a clear reddening and decline of the transient; see \citep{Hall2025sn}.

\paragraph{The two meter twins telescopes :} TTT~\citep{2024EPSC...17..631S}, sited at the Teide Observatory of the Instituto de Astrofísica de Canarias (IAC), started observations 2.9 days post $T_0$ of ZTF\,25abjmnps, in $g$, $r$, $i$ and $z$-bands. We detected the transient in $g$, $r$, $i$ and obtained upper limits in $z$. We used the STDPipe pipeline and its web interface, to perform force photometry and template subtraction~\citep{Karpov_2025}. For calibration, we used PS1 catalog~\citep{2016arXiv161205560C}. For subtraction templates we use Legacy Survey DR10 images~\citep{2019AJ....157..168D} for $g$, and $z$ images and PS1 for images in $i$-band, MegaCam for $r$ from the \href{https://www.skysurvey.cc/releases/}{UNIONS} survey. We performed the subtraction using HOTPANTS\citep{2015ascl.soft04004B} with convolution kernel size adjusted for every image individually based on the FWHMs of image and template. 
%As the object is located on top of the bright galaxy, and the passbands of the image and template are not exactly the same, the galaxy is not perfectly subtracted in most of our images, leading to subtraction residuals that might impact the photometry. 
%To account for that, we both reduced the aperture size (to reduce the contribution from biased background estimation inside it) and increased the error estimate from purely statistical one to also include this systematic error, by increasing the error bars by 0.1 mag at minimum fainter than 21.5 mag and 0.05 mag at minimum brighter than 21.5 mag   (typical value of measurement variations when changing the aperture size, thus roughly corresponding to the background bias).

\paragraph{CFHT MegaCam:} We acquired data using the MegaPrime camera mounted on the Canada French Hawaii Telescope, in the $g$, $r$, $i$, and $z$-band taken from 21st of August to 29th of August. Similarly to TTT, we used the STDPipe pipeline and its web interface, to perform force photometry and template subtraction~\citep{Karpov_2025}. 
%For subtraction templates we use Legacy Survey DR10 images~\citep{2019AJ....157..168D} for $g$, and $z$ images and PS1 for images in $i$-band and the prior observations from MegaCam taken in 2017 from the \href{https://www.skysurvey.cc/releases/}{UNIONS} survey.

\paragraph{Liverpool Telescope (LT):} The robotic 2m Liverpool Telescope \citep{2004SPIE.5489..679S} observed the location of the transient on several occasions using the IO:O optical imager in $g$, $r$, $i$, and $z$ bands.  Images were automatically processed by the LT data pipeline and images were subtracted using Pan-STARRS imaging as a template.  PSF photometry is performed on the subtracted images.

\paragraph{The Spectral Energy Distribution Machine (SEDM):} Mounted on the 60-inch Telescope at Palomar Observatory, SEDM \citep{sedm} took images in the $g$, $r$, and $i$-band throughout the duration of our campaign. Standard reduction techniques were applied to the data, and image subtraction revealed a clear excess at the location of the transient in multiple epochs. 

\paragraph{The Wide Field Infrared Camera (WIRC):} We acquired $J$- and $Ks$-band images with WIRC, mounted on the 200-inch Hale Telescope in Palomar observatory. We performed standard calibrations, subtracted a UKIRT template to the images in $J$-band, and used the 2MASS \citep{Skrutskie2006} catalog to calibrate the photometry. % add new WIRC template take n the 16th determine  taken on XX September, 2025. Our observations show XX.

\paragraph{The multi-object spectrometer for infra-red exploration (MOSFIRE):} We used MOSFIRE \cite{mosfire}, mounted in the Keck II telescope, to observe ZTF\,25abjmnps in the $Y$, $J$, $H$, and $Ks$-band. We used \texttt{mirar} to reduce the images and the Pan-STARRS and 2MASS catalog for photometric calibration. When available, we performed image subtraction against the Hubble Space Telescope image. Otherwise, we used a \texttt{galfit} \citep{galfit} to model the galaxy with an exponential and Sersic profile to subtract the galaxy light. 

\paragraph{APO} We observed with the Near-Infrared Camera \& Fabry-Perot Spectrometer \citep[NICFPS;][]{Hearty:04:NICFPS} mounted on the ARC 3.5\,m telescope at Apache Point Observatory.
We obtained dithered exposures in $Z$, $J$, $H$, and $Ks$.
We subtracted dark frames and constructed sky flats using custom Python routines (B.\ Bolin, priv. comm.) adapted from \citet{Weisenburger2017} and coadded the resulting images with SWarp \citep{2002ASPC..281..228B}.

\paragraph{The Low Resolution Imaging Spectrometer (LRIS):} 
We used LRIS \citep{lris}, mounted in the Keck I telescope, to acquire $g$ and $I$-band photometry of the ZTF\,25abjmnps. Observations started with a 30 s sequence, followed by a 300 s exposure. We used \texttt{lpipe} \citep{lpipe} to reduce the images and we photometrically calibrated the images against Pan-STARRS.

\paragraph{MMT} We collected optical photometry and spectroscopy with Binospec \citep{fabricant_binospec_2019} mounted on the 6.5\,m MMT telescope at the Fred Lawrence Whipple Observatory. Photometric observations were taken with the r, i, and z bands and reduced with the POTPyRI pipeline\footnote{\href{https://github.com/CIERA-Transients/POTPyRI}{https://github.com/CIERA-Transients/POTPyRI}}. The astrometry is calibrated using Gaia DR3 astrometric standard stars, and the flux measurements are calibrated against PS1. We realized the subtraction on the images from the galaxy similarly to TTT and CFH data analysis. Spectroscopic observations were taken with the 270 lines/mm grating, achieving R$\sim1340$ with coverage between 3820-9210~\AA. The spectroscopic reductions use \texttt{pypeit} \citep{prochaska_pypeit_2020} to automatically execute standard reduction procedures.

\paragraph{The Gemini Multi-Object Spectrograph (GMOS):}  With GMOS mounted on the Gemini-North telescope, we took slit spectrography on 2025-09-19 and imaged the location of ZTF\,25abjmnps over 6 epochs between 2025-08-20 and 2025-09-31 in $g$, $r$, $i$ and $z$-bands (PI Palmese; PI: O'Connor).  The images were reduced with the \textsc{DRAGONS} pipeline \citep{2019ASPC..523..321L} and difference imaging was conducted with SFFT \citep{hu_image_2022}. %using the deepest available archival templates which comprise CFHT Megacam in $r$-band, the Legacy Survey DR10 \citep{2019AJ....157..168D} for $g$ and $z$-bands and Pan-STARRS1 \citep[PS1;][]{2010SPIE.7733E..0EK} in $i$-band.  
Aperture photometry was conducted on the difference images using PS1 to calculate magnitude zeropoints. The spectra was reduced with pypeit, with telluric corrections done through atmospheric modeling \citep{prochaska_pypeit_2020}. More details in \citep{Hall2025sn}. 

\paragraph{Blanco} We observed with the Dark Energy Camera (DECam) instrument mounted on the Blanco 4m Telescope at the Cerro Tololo Inter-American Observatory (CTIO) using the $g$, $r$, and $z$ bands (PI: Palmese). The images were reduced with astrometric calibration against Gaia DR3 \citep{Gaia2021, 2021A&A...649A...2L, gaiaEDR3} and photometric calibration against PS1. We use SFFT \citep{hu_image_2022} for image subtraction against archival DECam images (Hu et al. in prep). We clearly resolve the rise in all filters; see \citealt{Hall2025sn} for more details.

\paragraph{Himalayan Chandra Telescope (HCT):} 
We observed the transient with the 2.0\,m Himalayan Chandra Telescope (HCT) at the Indian Astronomical Observatory (IAO), obtaining multiple exposures in the SDSS $r'$ and $i'$ filters starting on 9th September. Image subtraction was performed using the \texttt{ZOGY}-based \texttt{Python} pipeline \citep{2022MNRAS.516.4517K, 2016ApJ...830...27Z}, with CFHT/MegaCam templates for $r'$ and Pan-STARRS templates for $i'$. Basic reduction, astrometry, and point-spread-function (PSF) photometry followed the procedure described in \citet{2022AJ....164...90K}. The final magnitudes were calibrated against Pan-STARRS.

\section{Additional candidate vetting}
In light of the superkilonova model, we re-examined the ZTF data with looser selection cuts to identify any other young SNe that could potentially be linked to S250818k through the superkilonova model. Using the python package \texttt{emgwcave}\footnote{https://github.com/virajkaram/emgwcave}, we conducted an archival search for all candidates within the 95\% localization region of S250818k that were first detected by ZTF within 4 days of the GW trigger, have at least two ZTF detections, do not have host-galaxy redshifts (photometric or spectroscopic) outside the 3D GW localization, and do not coincide with cataloged AGN. We also reject sources that do not show any substantial photometric variations in their ZTF photometry to reject old SNe. In addition to ZTF\,25abjmnps, we find six sources that are candidate infant supernovae exploding around the time of GW trigger. Next, we discuss each of these six sources in some detail to assess whether any of them could be young stripped envelope supernovae consistent with the superkilonova model.

Two of these --- ZTF\,25abkoomo (SN\,2025vfa \citealt{2025vfa_disc, 2025vfa_class}) and ZTF25\,abjvflp (SN\,2025uxs \citealt{2025uxs_disc, 2025uxs_class}) --- have been spectroscopically classified as Type Ia SNe. Thus, we reject these two thermonuclear explosions in the superkilonova picture. 

ZTF\,25abkeuac (SN\,2025uic, \citealt{2025uic_disc,2025uic_class}) is a spectroscopically classified Type II SN. Since it is not a stripped envelope supernova, we consider it to be unrelated in the superkilonova model.  

ZTF\,18aawigkf (SN\,2025uso, \citealt{Chambers2025_2025uso, Sollerman2025_2025uso}, also reported in \citealt{Franz2025, Gillanders2025}) is a spectroscopically classified Type IIb SN. The first ZTF detection for this source is 3 days after the GW trigger, and we have ZTF non-detections of $g>21.77$, $r>21.83$ from MJD 60905.18 and 60905.23, respectively, corresponding to  (M$_{g/r}\approx-14$). We fit the ZTF light curve with a power law to constrain the explosion time. The joint fit to the g-band and r-band light curve gives an explosion time estimate that is 2.27$^{+0.31}_{-0.06}$ days after the gravitational wave trigger (error bars are 3$\sigma$). Since the supernova cannot explode after the kilonova in the superkilonova picture, we rule out association between SN\,2025uso and S250818k. 

ZTF\,25abjvfjh (AT\,2025wxt \citealt{2025wxt_disc}) is an unclassified transient that was first detected 0.15 hours after the GW trigger and brightened by 2 magnitudes in the next twenty days, before going into solar conjunction. This source has a photometric redshift of 0.134$\pm0.025$ from SDSS, which is only marginally consistent with the 3-$\sigma$ limits from the GW detection. While we cannot conclusively exclude this transient, it is likely too distant to be associated with the GW trigger.  

One other hostless transient, ZTF\,25abjmput (AT\,2025unk \citealt{2025unk_disc}) has limited photometric data that shows possible fading in its  brightness. However, the age of this source cannot be determined, as no constraining non-detections exist for this source before its first detection. This source is consistent with being a late-time fading supernova, and is likely unrelated to the superkilonova picture.

In addition to these six sources discussed above that were flagged in our re-analysis,  we also comment on two other sources --- AT2025uow and AT2025uxu --- reported by \cite{Franz2025}
as sources with higher rank than AT2025ulz. AT\,2025uxu was detected by ZTF and included in our candidate vetting, but was ruled out due to a photometric redshift of z$\approx$0.102 of its host galaxy from \cite{Saulder2023}. AT\,2025uow is too far south and was not covered by ZTF observations. However, as noted in \citealt{Franz2025}, there is an ATLAS pre-detection two days before the GW trigger which would rule it out in the superkilonova picture. 

In summary, our re-analysis indicates that ZTF\,25abjmnps (AT2025ulz) remains the only plausible candidate counterpart to S250818k in the superkilonova picture.

\clearpage

\begin{longtable*}{c | c | c | c| c }
\hline
ZTF Name & TNS Name & ra [deg] & dec [deg] & Rejection Reason\\
\hline
\endhead
\hline
ZTF25abjmlzh & / & 243.2441434 & 38.4684582 & AGN\\ 
ZTF25abjmmee & / & 239.2453486 & 40.1835372 & AGN\\ 
ZTF25abjmmiw & / & 239.3186678 & 35.6646666 & AGN\\ 
ZTF25abjmmnb & / & 248.6882847 & 41.0338104 & AGN\\ 
ZTF25abjmmny & / & 243.6915066 & 43.8616475 & AGN\\ 
ZTF25abjmmqa & / & 240.8577739 & 35.3408267 & AGN\\ 
ZTF25abjmneg & / & 240.0118714 & 32.4457522 & AGN\\ 
ZTF25abjmnjl & / & 247.6316529 & 38.5907847 & AGN\\ 
ZTF25abjmnjm & / & 246.9054209 & 38.5110897 & AGN\\ 
ZTF25abjmnlu & / & 243.4644548 & 38.5898236 & AGN\\ 
ZTF25abjmoij & / & 231.2047775 & 25.4980714 & AGN\\ 
ZTF25abjmoim & / & 230.7241176 & 25.436271 & AGN\\ 
ZTF25abjmopk & / & 232.5895058 & 23.1978009 & AGN\\ 
ZTF25abjmpbu & / & 235.0880589 & 35.0228726 & AGN\\ 
ZTF25abjmpcc & / & 238.613956 & 32.5973821 & AGN\\ 
ZTF25abjmpdk & / & 237.5400625 & 36.0923716 & AGN\\ 
ZTF25abjmpft & / & 235.9457219 & 32.8679695 & AGN\\ 
ZTF25abjmpls & / & 236.6196712 & 36.4053266 & AGN\\ 
ZTF25abjmpmy & / & 234.181787 & 30.6227189 & AGN\\ 
ZTF25abjmpsa & / & 238.5396497 & 29.9214224 & AGN\\ 
ZTF25abjmpuy & / & 245.7261016 & 40.5976814 & AGN\\ 
ZTF25abjmpyx & / & 246.8505705 & 40.8441978 & AGN\\ 
ZTF25abjmpzl & / & 249.1172467 & 40.5117299 & AGN\\ 
ZTF25abjmror & / & 247.9727637 & 39.6356484 & AGN\\ 
ZTF25abjmrps & / & 243.2863406 & 40.0484191 & AGN\\ 
ZTF25abjmrwg & / & 247.4124045 & 40.0957375 & AGN\\ 
ZTF25abjmsvl & / & 242.9030938 & 33.0607713 & AGN\\ 
ZTF25abjmszb & / & 241.8772998 & 32.0510167 & AGN\\ 
ZTF25abjmuox & / & 242.7361617 & 40.4205536 & AGN\\ 
ZTF25abjmuzq & / & 236.2768791 & 31.8180496 & AGN\\ 
ZTF25abjmvas & / & 237.2899228 & 33.1738036 & AGN\\ 
ZTF25abjmvlp & / & 251.3286507 & 43.5840673 & AGN\\ 
ZTF25abjmvng & / & 243.8545385 & 43.9292681 & AGN\\ 
ZTF25abjmvns & / & 250.0968232 & 42.4055574 & AGN\\ 
ZTF25abjmwig & / & 265.9694206 & 51.9391237 & AGN\\ 
ZTF25abjmwqt & / & 254.9181861 & 50.4702675 & AGN\\ 
ZTF25abjnaty & / & 247.6235338 & 40.4648626 & AGN\\ 
ZTF25abjvdhc & / & 246.0949246 & 38.9379507 & AGN\\ 
ZTF25abjvdsy & / & 248.127781 & 42.7333932 & AGN\\ 
ZTF25abjvehh & / & 239.611453 & 37.4682484 & AGN\\ 
ZTF25abjvelr & / & 235.8163818 & 35.2687048 & AGN\\ 
ZTF25abjvfcx & / & 237.5587453 & 28.829855 & AGN\\ 
ZTF25abjvfdl & / & 236.6740881 & 32.981276 & AGN\\ 
ZTF25abjvfqn & / & 233.5591153 & 24.857189 & AGN\\ 
ZTF25abjvgpb & / & 236.1395022 & 30.1447407 & AGN\\ 
ZTF25abjvift & / & 262.60477 & 49.9013771 & AGN\\ 
ZTF25abjvdlb & / & 247.6483527 & 38.7868027 & BOGUS\\ 
ZTF25abjvebe & / & 241.6307989 & 35.1487304 & BOGUS\\ 
ZTF25abjvekf & / & 234.7316111 & 34.8548818 & BOGUS\\ 
ZTF25abjvelp & / & 235.3035119 & 35.5632559 & BOGUS\\ 
ZTF25abjveur & / & 240.3990312 & 35.5260795 & BOGUS\\ 
ZTF25abjvfar & / & 242.4176028 & 35.4784746 & BOGUS\\ 
ZTF25abjvfjt & / & 235.8557562 & 24.3259834 & BOGUS\\ 
ZTF25abjvfqi & / & 234.0581773 & 24.3453163 & BOGUS\\ 
ZTF25abjmotz & AT 2025uzg & 233.7460861 & 24.8628222 & FAR - specz=0.229\\ 
ZTF25abjmmtx & AT 2025uzk & 239.9571177 & 33.5514903 & FAR - photz l95 = 0.139\\ 
ZTF25abjmmvx & AT 2025uzl & 241.6009779 & 37.7817471 & FAR - specz=0.200\\ 
ZTF25abjmooj & AT 2025uzo & 234.0232051 & 25.5989501 & FAR - photz l95=0.104\\ 
ZTF25abjmoqi & AT 2025uzp & 229.7612588 & 25.6648619 & FAR - photz l95 = 0.239\\ 
ZTF25abjmotq & AT 2025uzq & 232.9913306 & 25.2775372 & FAR - photz l95 = 0.24\\ 
ZTF25abjmtah & AT 2025uzs & 240.9929325 & 32.4011125 & FAR - photz l95 = 0.278\\ 
ZTF25abjmrzd & AT 2025uzt & 236.24615 & 35.7672837 & FAR - specz=0.282\\ 
ZTF25abjmoko & AT 2025uzv & 232.9373295 & 25.0276915 & FAR - photz l95 = 0.198\\ 
ZTF25abjmmqb & AT 2025uzw & 241.2051121 & 35.2970649 & FAR - photz l95 = 0.256\\ 
ZTF25abjmmbg & AT 2025uzx & 242.0437589 & 38.3652593 & FAR - photz l95 = 0.301\\ 
ZTF25abjmuvw & AT 2025uzy & 231.0306281 & 23.8804767 & FAR - photz l95 = 0.242\\ 
ZTF25abjmmpg & AT 2025uzz & 238.4060566 & 36.1216131 & FAR - photz l95 = 0.398\\ 
ZTF25abjmwzo & AT 2025vaa & 254.1144532 & 46.8250141 & FAR - photz l95 = 0.421\\ 
ZTF25abjmxrk & AT 2025vab & 241.8166578 & 36.9920751 & FAR - photz l95 = 0.104\\ 
ZTF25abjvdsl & AT 2025vac & 239.7555799 & 40.2065827 & FAR - photz l95 = 0.755\\ 
ZTF25abjvfkj & AT 2025vad & 237.2357957 & 28.6867656 & FAR - photz l95 = 0.334\\ 
ZTF25abjvfql & AT 2025vae & 234.2499963 & 24.4685611 & FAR - specz=0.206\\ 
ZTF25abjvddz & AT 2025vaf & 244.7486917 & 38.1097621 & FAR - photz l95 = 0.13\\ 
ZTF25abjvgcc & AT 2025vag & 232.3370831 & 24.4239568 & FAR - photz l95 = 0.614\\ 
ZTF25abjvfzu & AT 2025vah & 239.4819984 & 29.2197791 & FAR - photz l95 = 0.554\\ 
ZTF25abjmtkb & AT 2025vai & 242.0910463 & 32.7409275 & FAR - photz l95 = 0.175\\ 
ZTF25abjmlyq & AT 2025vaj & 245.1994494 & 42.0982548 & FAR - specz=0.229\\ 
ZTF25abjmman & AT 2025vak & 237.61834 & 38.5752129 & FAR - photz l95 = 0.135\\ 
ZTF25abjmlsw & AT 2025val & 245.4769086 & 40.0921368 & FAR - photz l95 = 0.198\\ 
ZTF25abjvgrv & AT 2025vam & 235.0854361 & 30.5154994 & FAR - photz l95 = 0.418\\ 
ZTF25abjmvte & AT 2025van & 247.9201141 & 40.5134422 & FAR - photz l95 = 0.173\\ 
ZTF25abjmujt & AT 2025var & 232.656017 & 28.8539315 & FAR - specz=0.800\\ 
ZTF25abjmtyf & AT 2025vas & 242.604702 & 36.9211029 & FAR - photz l95 = 0.115\\ 
ZTF25abjmoic & AT 2025vat & 233.0135327 & 27.5936511 & FAR - photz l95 = 0.608\\ 
ZTF25abjmova & AT 2025vau & 234.3891405 & 29.6584142 & FAR - specz=0.578\\ 
ZTF25abjmohh & AT 2025vax & 241.078763 & 32.3592959 & FAR - specz=0.319\\ 
ZTF25abjmogw & AT 2025vay & 230.7165663 & 24.8830003 & FAR - photz l95 = 0.183\\ 
ZTF25abjmnsq & AT 2025uzm & 234.263149 & 31.8666658 & FAR - photz l95 = 0.193\\ 
ZTF25abjmnuh &  
AT 2025uzn & 235.2293025 & 31.3957552 & FAR - photz l95 = 0.094\\ 
ZTF25abjmpck &  
AT 2025vav & 235.3139675 & 32.6537897 & FAR - photz l95 = 0.481\\ 
ZTF25abjmxii & AT 2025uzf & 255.1250966 & 49.6712643 & FAR - photz SDSS = 0.109  \\
%SLOW - $\delta g / \delta t = 0.54$ mag d$^{-1}$\\ 
ZTF25abjmlwp & AT 2025unj & 242.17952 & 40.0270559 & OLD - FP\\ 
ZTF25abjmtcr & AT 2025uzr & 241.442385 & 37.0062011 & OLD - FP\\ 
ZTF25abjmptv & AT 2025uzu & 244.2891574 & 41.7935382 & OLD - FP\\ 
ZTF25abjmxjc & AT 2025vao & 257.7599607 & 48.702757 & OLD - FP\\ 
ZTF25abjmput & AT 2025unk & 246.3894872 & 40.7303896 & SLOW - $\delta g / \delta t = 0.06$ mag d$^{-1}$\\ 
ZTF25abjmvvb & AT 2025unl & 250.1465103 & 46.743553 & SLOW - $\delta g / \delta t = -0.03$ mag d$^{-1}$\\ 
ZTF25abjmvwh & AT 2025unm & 247.0347148 & 42.0572803 & SLOW - $\delta g / \delta t = 0.07$ mag d$^{-1}$\\ 
ZTF25abjmmpa & AT 2025unn & 239.1654626 & 33.2561054 & SLOW - $\delta g / \delta t = -0.05$ mag d$^{-1}$\\ 
ZTF25abjmoef & AT 2025uno & 231.2551448 & 26.2795755 & SLOW - $\delta g / \delta t = 0.08$ mag d$^{-1}$\\ 
ZTF25abjmmps & AT 2025unp & 241.024028 & 35.6278231 & SLOW - $\delta g / \delta t = -0.01$ mag d$^{-1}$\\ 
ZTF25abjvhqg & AT 2025uqe & 257.9830966 & 49.1794007 & SLOW - $\delta g / \delta t = -0.01$ mag d$^{-1}$\\ 
ZTF25abjmoeu & AT 2025uzc & 235.9834181 & 25.1266572 & SLOW - $\delta g / \delta t = -0.13$ mag d$^{-1}$\\ 
ZTF25abjmvof & AT 2025uze & 306.3230356 & 67.3742981 & SLOW - $\delta g / \delta t = -0.10$ mag d$^{-1}$\\ 
ZTF25abjmntm & AT 2025uzh & 239.1767522 & 34.153721 & SLOW - $\delta g / \delta t = -0.04$ mag d$^{-1}$\\ 
ZTF25abjmnql & AT 2025uzi & 244.8820228 & 36.804871 & SLOW - $\delta g / \delta t = -0.20$ mag d$^{-1}$\\ 
ZTF25abjmotj & AT 2025uzj & 233.5968358 & 23.1898252 & SLOW - $\delta g / \delta t = -0.24$ mag d$^{-1}$\\ 
ZTF25abjmmub & / & 239.9616469 & 40.4055308 & STELLAR\\ 

\caption{All rejected ZTF candidate counterparts to S250818k, alongside their primary rejection reason. Candidates flagged as AGN are nuclear and have AGN-like WISE colours \citep{wise, 2005ApJ...631..163S}, are listed in Milliquas \citep{2023OJAp....6E..49F} or are classified as AGN using DESI spectrum \citep{2025arXiv250314745D}. Candidates flagged BOGUS were noted as subtraction residuals based on manual visual vetting. Candidates flagged as STELLAR have a point source counterpart in the reference image. AGN, STELLAR and BOGUS candidates were not reported to TNS. Candidates flagged as FAR are based on the host redshift (spectroscopic from DESI \citep{2025arXiv250314745D} or SDSS \citep{sdss_00}, or photometric from Legacy Survey DR9 \footnote{\url{https://www.legacysurvey.org/dr9}}) being inconsistent with the GW volume. Candidates flagged as OLD are rejected based on detections recovered by forced photometry on ZTF data that precede the GW trigger time. Candidates rejected as SLOW are based on photometric evolution being slower than 0.3 mag/day. The rejection reasons are not mutually exclusive, so for example, a FAR candidate could also be SLOW. Host galaxy spectra presented in \citealt{Hall2025desi, hall_at2025ulz_2025} more firmly reject 2025uzf, 2025vaa, 2025uzx, 2025vad, 2025vat, 2025unn, 2025unm, 2025unl, 2025unk, 2025vak, and 2025vah as too far and not consistent with S250818k.}
\label{tab:candidates}
\end{longtable*}
\clearpage

\begin{longtable*}{c | c | c | c| c | c | c}
\hline
Time [UT]&Phase&Mag [AB]&$\Delta_{m}$& Limiting Mag & Filter&Instrument\\
\hline
\endhead
\hline
2025-08-13 04:21:13 & -4.87 d & / & / & 21.2 & ztfg & ZTF\\ 
2025-08-13 06:45:28 & -4.77 d & / & / & 20.2 & ztfr & ZTF\\ 
2025-08-15 05:13:45 & -2.84 d & / & / & 21.0 & ztfg & ZTF\\ 
2025-08-15 06:46:21 & -2.77 d & / & / & 20.6 & ztfr & ZTF\\ 
2025-08-17 04:15:59 & -0.88 d & / & / & 21.1 & ztfg & ZTF\\ 
2025-08-17 05:11:56 & -0.84 d & / & / & 21.0 & ztfr & ZTF\\ 
2025-08-18 04:31:15 & 0.13 d & 21.0 & 0.1 & 21.6 & ztfg & ZTF\\ 
2025-08-18 04:41:35 & 0.14 d & 21.2 & 0.1 & 22.2 & ztfg & ZTF\\ 
2025-08-18 05:50:10 & 0.19 d & 21.3 & 0.1 & 21.9 & ztfr & ZTF\\ 
2025-08-18 06:48:37 & 0.23 d & 21.1 & 0.1 & 21.5 & ztfg & ZTF\\ 
2025-08-18 19:35:11 & 0.76 d & 21.3 & 0.1 & 23.9 & sdssg & FTW 3KK\\ 
2025-08-18 19:35:11 & 0.76 d & 21.5 & 0.1 & 23.1 & sdssi & FTW 3KK\\ 
2025-08-18 20:15:49 & 0.79 d & 21.5 & 0.1 & 23.8 & sdssg & FTW 3KK\\ 
2025-08-18 20:15:49 & 0.79 d & 21.4 & 0.1 & 23.1 & sdssi & FTW 3KK\\ 
2025-08-18 21:15:53 & 0.83 d & 21.5 & 0.1 & 22.1 & sdssz & FTW 3KK\\ 
2025-08-18 21:15:53 & 0.83 d & 21.7 & 0.1 & 23.5 & sdssr & FTW 3KK\\ 
2025-08-18 21:15:53 & 0.83 d & 21.8 & 0.2 & 22.7 & sdssz & FTW 3KK\\ 
2025-08-18 21:51:31 & 0.86 d & 21.4 & 0.1 & 23.7 & sdssg & FTW 3KK\\ 
2025-08-18 21:51:31 & 0.86 d & 21.6 & 0.1 & 23.0 & sdssi & FTW 3KK\\ 
2025-08-19 03:40:55 & 1.10 d & 21.6 & 0.1 & 22.0 & sdssg & SEDM\\ 
2025-08-19 03:47:23 & 1.10 d & 21.7 & 0.4 & 21.0 & sdssr & SEDM\\ 
2025-08-19 04:20:15 & 1.13 d & 21.4 & 0.2 & 21.6 & ztfg & ZTF\\ 
2025-08-19 19:22:06 & 1.75 d & 22.0 & 0.2 & 22.2 & sdssz & FTW 3KK\\ 
2025-08-19 19:22:06 & 1.75 d & 22.1 & 0.1 & 23.6 & sdssr & FTW 3KK\\ 
2025-08-19 19:22:06 & 1.75 d & 21.9 & 0.2 & 22.8 & sdssz & FTW 3KK\\ 
2025-08-19 20:09:46 & 1.78 d & 22.1 & 0.1 & 24.0 & sdssg & FTW 3KK\\ 
2025-08-19 20:09:46 & 1.78 d & 21.7 & 0.1 & 23.1 & sdssi & FTW 3KK\\ 
2025-08-19 21:49:25 & 1.85 d & 22.2 & 0.1 & 23.1 & sdssg & IOO\\ 
2025-08-19 21:57:46 & 1.86 d & 22.1 & 0.1 & 23.1 & sdssr & IOO\\ 
2025-08-19 22:06:05 & 1.87 d & 21.6 & 0.1 & 23.1 & sdssi & IOO\\ 
2025-08-20 05:31:06 & 2.17 d & 21.6 & 0.1 & 24.0 & sdssz & GMOS\\ 
2025-08-20 05:50:25 & 2.19 d & 21.7 & 0.1 & 24.0 & sdssi & GMOS\\ 
2025-08-20 05:58:23 & 2.19 d & 22.5 & 0.1 & 25.2 & sdssr & GMOS\\ 
2025-08-20 06:07:02 & 2.20 d & 22.7 & 0.1 & 25.7 & sdssg & GMOS\\ 
2025-08-20 07:04:54 & 2.24 d & 22.6 & 0.2 & 24.8 & sdssg & LRIS\\ 
2025-08-20 22:05:19 & 2.86 d & 22.3 & 0.3 & 22.2 & sdssi & TTT\\ 
2025-08-20 22:26:00 & 2.88 d & 22.9 & 0.3 & 23.0 & sdssg & IOO\\ 
2025-08-20 22:45:51 & 2.89 d & 22.8 & 0.2 & 23.1 & sdssr & IOO\\ 
2025-08-21 03:18:17 & 3.08 d & / & / & 22.5 & 2massj & WIRC\\ 
2025-08-21 06:03:57 & 3.20 d & 22.6 & 0.1 & 24.5 & sdssz & GMOS\\ 
2025-08-21 06:08:39 & 3.20 d & 23.1 & 0.2 & 24.3 & sdssg & MegaCam\\ 
2025-08-21 06:52:45 & 3.23 d & 22.9 & 0.2 & 23.4 & sdssr & MegaCam\\ 
2025-08-21 07:45:49 & 3.27 d & 23.4 & 0.3 & 23.9 & sdssg & LRIS\\ 
2025-08-21 08:06:07 & 3.28 d & 23.1 & 0.1 & 24.5 & sdssg & GMOS\\ 
2025-08-22 04:52:43 & 4.15 d & 22.1 & 0.1 & 24.5 & sdssi & Binospec\\ 
2025-08-22 05:30:58 & 4.17 d & 22.5 & 0.1 & 24.0 & sdssz & GMOS\\ 
2025-08-22 05:51:00 & 4.19 d & 23.1 & 0.1 & 25.3 & sdssg & MegaCam\\ 
2025-08-22 06:08:37 & 4.20 d & 23.0 & 0.1 & 25.0 & sdssr & GMOS\\ 
2025-08-22 06:13:25 & 4.20 d & 22.1 & 0.1 & 25.0 & sdssi & MegaCam\\ 
2025-08-22 06:14:24 & 4.20 d & 23.1 & 0.2 & 24.0 & 2massh & MOSFIRE\\ 
2025-08-22 06:14:24 & 4.20 d & 22.9 & 0.2 & 24.0 & 2massj & MOSFIRE\\ 
2025-08-22 06:14:24 & 4.20 d & 23.3 & 0.3 & 24.0 & 2massks & MOSFIRE\\ 
2025-08-22 06:19:21 & 4.21 d & 23.0 & 0.1 & 25.0 & sdssg & GMOS\\ 
2025-08-22 06:42:32 & 4.22 d & 23.2 & 0.1 & 24.7 & sdssr & MegaCam\\ 
2025-08-22 20:32:43 & 4.80 d & 22.9 & 0.1 & / & F110W & HST (GCN 41506, \citealt{Yang2025})\\ 
2025-08-22 20:32:43 & 4.80 d & 22.8 & 0.3 & / & F160W & HST (GCN 41506, \citealt{Yang2025})\\ 
2025-08-23 03:11:31 & 5.08 d & / & / & 21.8 & 2massj & WIRC\\ 
2025-08-23 05:42:10 & 5.18 d & 23.2 & 0.1 & 24.5 & sdssg & GMOS\\ 
2025-08-23 07:23:24 & 5.25 d & 23.0 & 0.1 & 25.4 & sdssr & MegaCam\\ 
2025-08-23 07:43:36 & 5.27 d & 23.1 & 0.1 & 25.3 & sdssg & MegaCam\\ 
2025-08-23 08:06:03 & 5.28 d & 22.0 & 0.1 & 23.8 & sdssi & MegaCam\\ 
2025-08-24 07:44:41 & 6.27 d & 22.7 & 0.1 & 25.4 & sdssr & MegaCam\\ 
2025-08-24 08:12:03 & 6.29 d & 23.1 & 0.1 & 25.2 & sdssg & MegaCam\\ 
2025-08-24 19:33:24 & 6.76 d & 23.0 & 0.1 & 24.1 & sdssg & FTW 3KK\\ 
2025-08-24 19:33:24 & 6.76 d & 21.9 & 0.1 & 23.2 & sdssi & FTW 3KK\\ 
2025-08-24 21:50:38 & 6.85 d & 23.1 & 0.1 & 24.9 & sdssg & TTT\\ 
2025-08-24 21:55:50 & 6.86 d & 22.7 & 0.1 & 23.6 & sdssr & TTT\\ 
2025-08-24 22:11:28 & 6.87 d & 21.5 & 0.1 & 23.1 & sdssi & TTT\\ 
2025-08-25 05:30:27 & 7.17 d & 22.5 & 0.1 & 24.0 & sdssz & GMOS\\ 
2025-08-25 05:35:50 & 7.18 d & 21.8 & 0.1 & 24.0 & sdssi & GMOS\\ 
2025-08-25 05:41:09 & 7.18 d & 22.5 & 0.1 & 24.0 & sdssr & GMOS\\ 
2025-08-25 07:55:57 & 7.27 d & 22.4 & 0.1 & 24.7 & sdssr & MegaCam\\ 
2025-08-25 08:23:03 & 7.29 d & 22.7 & 0.1 & 24.9 & sdssg & MegaCam\\ 
2025-08-25 19:38:59 & 7.76 d & 22.8 & 0.1 & 24.0 & sdssg & FTW 3KK\\ 
2025-08-25 19:38:59 & 7.76 d & 22.7 & 0.1 & 24.3 & sdssg & FTW 3KK\\ 
2025-08-25 19:38:59 & 7.76 d & 21.6 & 0.1 & 23.2 & sdssi & FTW 3KK\\ 
2025-08-25 20:14:51 & 7.79 d & 22.2 & 0.1 & 23.6 & sdssr & FTW 3KK\\ 
2025-08-25 20:14:51 & 7.79 d & 22.2 & 0.1 & 23.9 & sdssr & FTW 3KK\\ 
2025-08-25 20:14:51 & 7.79 d & 22.3 & 0.2 & 23.0 & sdssz & FTW 3KK\\ 
2025-08-25 20:14:51 & 7.79 d & 22.0 & 0.1 & 22.6 & sdssz & FTW 3KK\\ 
2025-08-26 06:25:18 & 8.21 d & 22.6 & 0.1 & 25.5 & sdssg & MegaCam\\ 
2025-08-26 06:47:54 & 8.23 d & 21.4 & 0.1 & 23.4 & sdssi & MegaCam\\ 
2025-08-26 19:22:42 & 8.75 d & 21.5 & 0.2 & 22.4 & sdssi & FTW 3KK\\ 
2025-08-26 22:08:46 & 8.87 d & 21.9 & 0.1 & 23.8 & sdssr & TTT\\ 
2025-08-27 03:32:32 & 9.09 d & 21.2 & 0.2 & 21.4 & ztfi & ZTF\\ 
2025-08-27 05:59:34 & 9.19 d & / & / & 20.7 & 2massj & WIRC\\ 
2025-08-27 19:35:10 & 9.76 d & 22.4 & 0.2 & 23.2 & sdssg & FTW 3KK\\ 
2025-08-27 19:35:10 & 9.76 d & 21.2 & 0.1 & 22.7 & sdssi & FTW 3KK\\ 
2025-08-27 21:17:04 & 9.83 d & 22.4 & 0.1 & 23.6 & sdssg & IOO\\ 
2025-08-27 22:44:29 & 9.89 d & 21.4 & 0.1 & 23.4 & sdssi & IOO\\ 
2025-08-28 21:01:24 & 10.82 d & 21.3 & 0.1 & 23.8 & sdssr & IOO\\ 
2025-08-28 21:32:04 & 10.84 d & 21.1 & 0.1 & 23.5 & sdssi & IOO\\ 
2025-08-28 23:29:39 & 10.92 d & 21.5 & 0.1 & 21.8 & desz & DECam\\ 
2025-08-28 23:33:39 & 10.93 d & 21.5 & 0.1 & 22.8 & desr & DECam\\ 
2025-08-28 23:36:29 & 10.93 d & 21.9 & 0.1 & 23.1 & desg & DECam\\ 
2025-08-29 03:38:16 & 11.10 d & 21.4 & 0.2 & 21.6 & ztfg & ZTF\\ 
2025-08-29 03:58:05 & 11.11 d & 21.0 & 0.1 & 21.6 & ztfi & ZTF\\ 
2025-08-29 06:50:21 & 11.23 d & 21.7 & 0.1 & 25.2 & sdssg & MegaCam\\ 
2025-08-29 07:09:12 & 11.24 d & 21.2 & 0.1 & 25.2 & sdssr & MegaCam\\ 
2025-08-29 23:04:29 & 11.91 d & 21.2 & 0.2 & 21.7 & sdssi & FTW 3KK\\ 
2025-08-30 03:08:46 & 12.08 d & 21.4 & 0.3 & 21.3 & ztfg & ZTF\\ 
2025-08-30 03:28:27 & 12.09 d & 20.7 & 0.1 & 21.6 & ztfi & ZTF\\ 
2025-08-30 04:41:29 & 12.14 d & 21.2 & 0.1 & 20.4 & sdssz & Binospec\\ 
2025-08-30 04:59:17 & 12.15 d & 20.9 & 0.1 & 24.1 & sdssi & Binospec\\ 
2025-08-30 19:32:15 & 12.76 d & 21.9 & 0.1 & 23.8 & sdssg & FTW 3KK\\ 
2025-08-30 19:32:15 & 12.76 d & 21.0 & 0.1 & 23.0 & sdssi & FTW 3KK\\ 
2025-08-30 20:10:44 & 12.79 d & 21.4 & 0.1 & 23.4 & sdssr & FTW 3KK\\ 
2025-08-30 20:10:44 & 12.79 d & 21.1 & 0.1 & 22.4 & sdssz & FTW 3KK\\ 
2025-08-30 20:10:45 & 12.79 d & 21.4 & 0.2 & 21.8 & sdssz & FTW 3KK\\ 
2025-08-31 03:14:37 & 13.08 d & 21.3 & 0.1 & 21.8 & ztfr & ZTF\\ 
2025-08-31 03:27:05 & 13.09 d & 21.0 & 0.1 & 21.6 & ztfi & ZTF\\ 
2025-08-31 19:28:17 & 13.76 d & 21.4 & 0.1 & 23.8 & sdssr & FTW 3KK\\ 
2025-08-31 19:28:17 & 13.76 d & 21.0 & 0.1 & 23.0 & sdssi & FTW 3KK\\ 
2025-08-31 20:04:02 & 13.78 d & 21.5 & 0.2 & 22.5 & sdssz & FTW 3KK\\ 
2025-08-31 21:25:25 & 13.84 d & 21.2 & 0.1 & 23.6 & sdssr & TTT\\ 
2025-09-01 03:04:24 & 14.07 d & 21.5 & 0.3 & 21.2 & ztfg & ZTF\\ 
2025-09-01 03:23:11 & 14.09 d & 21.5 & 0.2 & 20.9 & sdssg & SEDM\\ 
2025-09-01 03:29:23 & 14.09 d & 21.4 & 0.1 & 21.9 & sdssr & SEDM\\ 
2025-09-01 03:29:23 & 14.09 d & 20.9 & 0.1 & 21.5 & ztfi & ZTF\\ 
2025-09-01 03:35:35 & 14.09 d & 21.1 & 0.3 & 20.7 & sdssi & SEDM\\ 
2025-09-02 03:35:01 & 15.09 d & 21.1 & 0.1 & 21.9 & sdssr & SEDM\\ 
2025-09-02 03:48:58 & 15.10 d & 20.5 & 0.1 & 21.6 & sdssi & SEDM\\ 
2025-09-02 03:48:58 & 15.10 d & 20.7 & 0.2 & 20.8 & sdssi & SEDM\\ 
2025-09-03 02:58:02 & 16.07 d & 21.5 & 0.3 & 21.0 & ztfg & ZTF\\ 
2025-09-03 03:12:40 & 16.08 d & 20.4 & 0.1 & 21.4 & ztfi & ZTF\\ 
2025-09-03 04:08:03 & 16.12 d & 21.3 & 0.2 & 21.4 & sdssg & SEDM\\ 
2025-09-03 04:11:55 & 16.12 d & 21.1 & 0.2 & 21.2 & sdssr & SEDM\\ 
2025-09-03 04:15:47 & 16.12 d & 20.6 & 0.1 & 21.1 & sdssi & SEDM\\ 
2025-09-03 04:15:47 & 16.12 d & 20.9 & 0.3 & 20.7 & sdssi & SEDM\\ 
2025-09-03 19:11:43 & 16.74 d & 21.7 & 0.1 & 23.4 & sdssg & FTW 3KK\\ 
2025-09-03 19:11:43 & 16.74 d & 20.8 & 0.1 & 23.0 & sdssi & FTW 3KK\\ 
2025-09-03 19:29:50 & 16.76 d & 21.3 & 0.1 & 22.5 & sdssz & FTW 3KK\\ 
2025-09-03 19:29:50 & 16.76 d & 21.2 & 0.1 & 23.4 & sdssr & FTW 3KK\\ 
2025-09-04 03:23:33 & 17.09 d & 20.7 & 0.1 & 21.3 & ztfi & ZTF\\ 
2025-09-05 19:20:12 & 18.75 d & 21.1 & 0.2 & 21.9 & sdssr & FTW 3KK\\ 
2025-09-05 19:38:28 & 18.76 d & 20.7 & 0.3 & 20.9 & sdssi & FTW 3KK\\ 
2025-09-06 03:15:54 & 19.08 d & 21.0 & 0.2 & 21.3 & sdssr & SEDM\\ 
2025-09-06 03:18:54 & 19.08 d & 20.6 & 0.1 & 21.5 & ztfi & ZTF\\ 
2025-09-06 03:22:27 & 19.08 d & 21.8 & 0.4 & 20.5 & sdssg & SEDM\\ 
2025-09-06 03:28:58 & 19.09 d & 20.3 & 0.1 & 21.2 & sdssi & SEDM\\ 
2025-09-06 03:28:58 & 19.09 d & 20.8 & 0.2 & 21.0 & sdssi & SEDM\\ 
2025-09-06 18:59:57 & 19.74 d & 21.0 & 0.1 & 23.1 & sdssr & FTW 3KK\\ 
2025-09-06 18:59:57 & 19.74 d & 21.1 & 0.1 & 22.5 & sdssz & FTW 3KK\\ 
2025-09-06 18:59:57 & 19.74 d & 20.9 & 0.1 & 22.0 & sdssz & FTW 3KK\\ 
2025-09-06 19:37:39 & 19.76 d & 21.8 & 0.2 & 22.3 & sdssg & FTW 3KK\\ 
2025-09-06 19:37:39 & 19.76 d & 20.8 & 0.1 & 22.2 & sdssi & FTW 3KK\\ 
2025-09-07 18:58:16 & 20.73 d & 21.7 & 0.1 & 23.7 & sdssg & FTW 3KK\\ 
2025-09-07 18:58:16 & 20.73 d & 20.6 & 0.1 & 23.0 & sdssi & FTW 3KK\\ 
2025-09-07 19:34:06 & 20.76 d & 21.1 & 0.1 & 22.3 & sdssz & FTW 3KK\\ 
2025-09-07 19:34:06 & 20.76 d & 21.1 & 0.1 & 23.3 & sdssr & FTW 3KK\\ 
2025-09-07 19:34:06 & 20.76 d & 21.0 & 0.1 & 22.7 & sdssz & FTW 3KK\\ 
2025-09-08 03:15:59 & 21.08 d & 20.6 & 0.1 & 21.2 & ztfi & ZTF\\ 
2025-09-09 03:14:34 & 22.08 d & 20.5 & 0.1 & 21.0 & ztfi & ZTF\\ 
2025-09-09 03:14:34 & 22.08 d & 20.5 & 0.1 & 21.0 & ztfi & ZTF\\ 
2025-09-09 03:20:15 & 22.08 d & 20.5 & 0.2 & 20.5 & sdssi & SEDM\\ 
2025-09-11 03:13:03 & 24.08 d & 21.2 & 0.2 & 20.8 & sdssr & SEDM\\ 
2025-09-11 03:16:55 & 24.08 d & 20.9 & 0.2 & 20.9 & sdssi & SEDM\\ 
2025-09-11 16:04:33 & 24.61 d & 20.8 & 0.1 & 21.9 & sdssr & HCT imager\\ 
2025-09-11 20:30:15 & 24.80 d & 21.0 & 0.2 & 23.8 & sdssr & TTT\\ 
2025-09-11 20:34:43 & 24.80 d & 22.3 & 0.1 & 24.3 & sdssg & TTT\\ 
2025-09-11 20:39:05 & 24.80 d & 20.9 & 0.1 & 23.2 & sdssi & TTT\\ 
2025-09-12 02:57:46 & 25.07 d & 21.0 & 0.1 & 21.9 & ztfr & ZTF\\ 
2025-09-12 03:10:19 & 25.08 d & 20.8 & 0.1 & 21.5 & ztfi & ZTF\\ 
2025-09-13 02:55:48 & 26.07 d & 21.1 & 0.1 & 21.7 & ztfr & ZTF\\ 
2025-09-13 03:08:19 & 26.08 d & 20.9 & 0.2 & 21.2 & ztfi & ZTF\\ 
2025-09-13 20:43:54 & 26.81 d & 21.1 & 0.1 & 22.9 & sdssr & IOO\\ 
2025-09-13 20:51:01 & 26.81 d & 21.0 & 0.1 & 22.5 & sdssi & IOO\\ 
2025-09-14 15:43:43 & 27.60 d & 21.1 & 0.1 & 22.4 & sdssr & HCT imager\\ 
2025-09-14 16:10:16 & 27.62 d & 20.8 & 0.1 & 21.2 & sdssi & HCT imager\\ 
2025-09-14 18:53:15 & 27.73 d & 21.7 & 0.1 & 23.6 & sdssr & FTW 3KK\\ 
2025-09-14 18:53:15 & 27.73 d & 21.4 & 0.1 & 22.5 & sdssz & FTW 3KK\\ 
2025-09-14 19:29:13 & 27.76 d & 21.0 & 0.2 & 21.5 & sdssi & FTW 3KK\\ 
2025-09-15 03:11:19 & 28.08 d & 20.9 & 0.3 & 20.6 & sdssi & SEDM\\ 
2025-09-16 20:20:50 & 29.79 d & 22.8 & 0.1 & 23.9 & sdssg & TTT\\ 
2025-09-16 20:31:20 & 29.80 d & 21.6 & 0.1 & 23.3 & sdssr & TTT\\ 
2025-09-18 18:17:11 & 31.71 d & 21.8 & 0.1 & 23.3 & sdssr & FTW 3KK\\ 
2025-09-18 18:17:11 & 31.71 d & 21.3 & 0.2 & 21.9 & sdssz & FTW 3KK\\ 
2025-09-18 19:03:40 & 31.74 d & 23.2 & 0.2 & 24.0 & sdssg & FTW 3KK\\ 
2025-09-18 19:03:40 & 31.74 d & 21.2 & 0.1 & 22.9 & sdssi & FTW 3KK\\ 
2025-09-19 18:27:59 & 32.71 d & 22.0 & 0.1 & 23.5 & sdssr & FTW 3KK\\ 
2025-09-19 19:03:55 & 32.74 d & 21.1 & 0.1 & 23.0 & sdssi & FTW 3KK\\ 
2025-09-20 05:08:23 & 33.16 d & 21.3 & 0.1 & 23.3 & sdssz & GMOS\\ 
2025-09-20 05:19:02 & 33.17 d & 22.4 & 0.1 & 24.4 & sdssr & GMOS\\ 
2025-09-20 05:27:44 & 33.17 d & 22.8 & 0.1 & 24.8 & sdssg & GMOS\\ 
2025-09-20 18:42:36 & 33.72 d & 22.1 & 0.1 & 23.1 & sdssr & FTW 3KK\\ 
2025-09-20 18:42:36 & 33.72 d & 21.5 & 0.3 & 21.7 & sdssz & FTW 3KK\\ 
2025-09-20 19:00:54 & 33.74 d & 21.2 & 0.1 & 22.9 & sdssi & FTW 3KK\\ 
2025-09-20 19:00:54 & 33.74 d & 23.3 & 0.2 & 23.9 & sdssg & FTW 3KK\\ 
2025-09-21 03:25:08 & 34.09 d & 23.1 & 0.1 & 25.0 & sdssg & Binospec\\ 
2025-09-21 18:40:47 & 34.72 d & 21.3 & 0.2 & 22.1 & sdssi & FTW 3KK\\ 
2025-09-23 05:14:12 & 36.16 d & 21.7 & 0.1 & 23.4 & sdssz & GMOS\\ 
2025-09-23 05:25:04 & 36.17 d & 21.7 & 0.1 & 23.2 & sdssi & GMOS\\ 
2025-09-23 05:33:44 & 36.18 d & 22.1 & 0.1 & 24.6 & sdssr & GMOS\\ 
2025-09-23 05:42:26 & 36.18 d & 22.9 & 0.1 & 24.6 & sdssg & GMOS\\ 
\caption{Photometric observations of AT 2025ulz. No magnitude is given in the case of non-detections or marginal detections below the nominal image depth, with the upper limit being given by the 5$\sigma$ limiting magnitude. Phase is given relative to the time of merger time.}
\label{tab:photometry}
\end{longtable*}

\end{document}